\documentclass{emulateapj}

\slugcomment{Draft version \today}

\shorttitle{Fundamental parameters and dust production in 47 Tuc}
\shortauthors{McDonald et al.}

\begin{document}

\title{Fundamental parameters, integrated RGB mass loss and dust production in the Galactic globular cluster 47 Tucanae}

\author{I.~McDonald\altaffilmark{1}, M.~L.~Boyer\altaffilmark{2}, J.~Th.~van Loon\altaffilmark{3}, A.~A.~Zijlstra\altaffilmark{1}, J.~L.~Hora\altaffilmark{4}, B.~Babler\altaffilmark{5}, M.~Block\altaffilmark{6}, K.~Gordon\altaffilmark{2}, M.~Meade\altaffilmark{5}, M.~Meixner\altaffilmark{2}, K.~Misselt\altaffilmark{6}, T.~Robitaille\altaffilmark{4}, M.~Sewi{\l}o\altaffilmark{2}, B.~Shiao\altaffilmark{2}, B.~Whitney\altaffilmark{2}}
\altaffiltext{1}{Jodrell Bank Centre for Astrophysics, Alan Turing Building, Manchester, M13 9PL, UK; iain.mcdonald-2@manchester.ac.uk}
\altaffiltext{2}{STScI, 3700 San Martin Drive, Baltimore, MD 21218, USA}
\altaffiltext{3}{Lennard-Jones Laboratories, Keele University, ST5 5BG, UK}
\altaffiltext{4}{Harvard-Smithsonian Center for Astrophysics, 60 Garden Street, MS 65, Cambridge, MA 02138-1516, USA}
\altaffiltext{5}{Department of Astronomy, University of Wisconsin, Madison, 475 North Charter Street, Madison, WI 53706-1582, USA}
\altaffiltext{6}{Steward Observatory, University of Arizona, 933 North Cherry Avenue, Tuscon, AZ 85721, USA}

\begin{abstract}
Fundamental parameters and time-evolution of mass loss are investigated for post-main-sequence stars in the Galactic globular cluster 47 Tucanae (NGC 104). This is accomplished by fitting spectral energy distributions (SEDs) to existing optical and infrared photometry and spectroscopy, to produce a true Hertzsprung--Russell diagram. We confirm the cluster's distance as $d = 4611 ^{+213}_{-200}$ pc and age as 12 $\pm$ 1 Gyr. Horizontal branch models appear to confirm that no more RGB mass loss occurs in 47 Tuc than in the more-metal-poor $\omega$ Centauri, though difficulties arise due to inconsistencies between the models. Using our SEDs, we identify those stars which exhibit infrared excess, finding excess only among the brightest giants: dusty mass loss begins at a luminosity of $\sim 1000$ L$_\odot$, becoming ubiquitous above $L = 2000$ L$_\odot$. Recent claims of dust production around lower-luminosity giants cannot be reproduced, despite using the same archival \emph{Spitzer} imagery.
\end{abstract}



\section{Introduction}
\label{IntroSect}

Circumstellar dust production, and the variation of its occurrence with a star's fundamental parameters, must be understood if we are to gain insight into both Galactic ecology and the history of the chemical enrichment of the Universe. The dusty winds of asymptotic giant branch (AGB) stars dominate the production of interstellar dust at all redshifts at which AGB stars are observed, and have therefore determined the chemical enrichment of Population I stars, including the Sun and Solar System \citep{Gehrz89,Sedlmayr94,Zinner03,VSBA09}. The integrated mass loss of RGB stars determines the envelope mass of a star leaving the RGB tip. It consequently determines its position on the horizontal branch (HB) and appears to be the major factor causing the `second parameter' (after metallicity) required to define HB morphology (e.g.\ \citealt{Rood73,Catelan00}). Mass loss on the asymptotic giant branch eventually ejects the star's entire hydrogen envelope, creating a post-AGB star and (perhaps) a planetary nebula (PN).

A number of factors hamper the quantitative examination of dust production. These including difficulties in determining outflow velocities in faint, metal-poor stars; the difficulty in determining grain density (or porosity), grain size and grain shape; the difficulty in identifying the mineralogy of grains being produced; even the difficulty in determining whether infrared excess is present at all. Open questions include how instantaneous and integrated RGB mass loss varies as a function of initial mass and metallicity \citep{FPB97,LC99,Catelan00}; whether dust can form efficiently around RGB stars; at what evolutionary stage dust production begins; and whether radiation pressure on oxygen-rich dust grains is sufficient to drive a wind, especially at low metallicity (e.g.\ \citealt{Lewis89}), or whether chromospheric (magneto-acoustic) driving or pulsation may also be important \citep{HM80,DHA84,Woitke06b,MvL07}.

Globular clusters provide an excellent laboratory in which we can examine these questions. They typically host a single or dominant population of stars at identical ages and metallicities, with sufficient number to be statistically useful. Comparisons within globular clusters therefore probe variations with evolution, while comparisons between globular clusters probe variations with metallicity and age. To provide a proper comparison, it is vital to compare fundamental parameters of stars. These can be determined by comparing spectral energy distributions (SEDs) to synthetic spectra from stellar atmosphere models. This approach has the advantage of allowing one to identify infrared excess, characteristic of dust production, in the SED.

In this study, we examine the fundamental parameters of stars in the globular cluster 47 Tucanae (NGC 104), one of the most-massive (6--9 $\times 10^5$ M$_\odot$; \citealt{SR75,MM85,MSS91}) and nearest ($\approx$4500 pc; \citealt{Harris96,PSvWK02,MAM06,SHO+07}) Galactic globular clusters, with a metallicity of [Fe/H] $\approx$ --0.7 \citep{Harris96,CG97,MB08,WCMvL10} and interstellar reddening of $E(B-V) = 0.04$ mag \citep{Harris96}. Throughout this paper, we will refer to two similar studies: \citet{MvLD+09}, which covers the cluster $\omega$ Centauri (NGC 5139; hereafter Paper I); and \citet{BMvL+09}, which covers the cluster NGC 362 (hereafter Paper II). 

Dust production in 47 Tuc is of particular interest due to the study of \citet{ORFF+07}. This study has claimed observational evidence of dust production along the entire RGB, in contrast to the observations of other clusters (e.g.\ $\omega$ Cen; Paper I). If this claim can be confirmed, it significantly increases the number of stars we know of that are returning dust to the interstellar medium. This would be particularly important at high ($z \sim 6$) redshift, where AGB stars remain the dominant dust producers, but cannot account for a galaxy's entire dust budget \citep{VSBA09}. Such claims must therefore be taken seriously and examined carefully. An analysis of the Origlia et al.\ claim was undertaken by \citet{BvLM+10}, who raise concern that apparently-red stellar colours may be a result of stellar blending in the densely populated cluster core ($\sim$50\,000 M$_\odot$ pc$^{-3}$ --- \citealt{GF95}) and data artifacts due to saturation of the brightest stars. \citet{ORF+10} defended their original hypothesis, suggesting that this dust is too warm to present considerable reddening of the [3.6]--[8] colour, but does present considerable excess in $K_{s}$--[8]. One of our aims herein must therefore be to examine this issue in more depth.

This paper follows a similar approach to Paper I. Section \ref{SectData} details the photometric data we use, its sources and its reduction; determination of stellar parameters for the cluster's stars by fitting their spectral energy distributions (SEDs); and the creation of a Hertzsprung--Russell diagram (HRD). Section \ref{SectIsos} covers the fitting of stellar isochrones to the HRD. In Section \ref{SectMdot}, we derive mid-infrared (mid-IR) excesses for our stars, based on our SEDs, and use these to determine which stars are dusty and compare them to those in \citet{ORFF+07}. The mass-loss rates and dust compositions of these stars are analysed in an accompanying work (Paper IV).


\section{Data reduction}
\label{SectData}

\subsection{The input data}
\label{InputSect}

\begin{center}
\begin{table*}
\caption{Input photometric data and their coverage of the cluster.}
\label{CoverageTable}
\begin{tabular}{lll}
    \hline \hline
Wavelength	& Source		& Coverage	\\
    \hline
$UBVi$			& MCPS; \citet{ZHT+02}	& West of cluster, avoids core \\
$UBVI_{C}$		& \citet{Stetson00}	& $\sim 20 ^\prime \times 20 ^\prime$ around core \\
$UBVI_{J}JHK_{s}$		& \citet{SHO+07}	& Northern two-thirds of core \\
$JHK_{s}$			& 2MASS; \citet{SCS+06}	& Entire cluster \\
3.6--8 $\mu$m$^1$	& SAGE-SMC; (Gordon et al., in prep.) & South-west of cluster, not including core\\
3.6--8 $\mu$m		& \citet{ORFF+07}$^2$	& Cluster core \& immediate west \& east \\
24 $\mu$m$^3$		& \citet{BBW+09}	& Cluster core \& immediate south-west \\
    \hline
\multicolumn{3}{p{0.9\textwidth}}{$^1$\emph{Spitzer} IRAC at 3.6, 4.5, 5.8 and 8 $\mu$m; \citet{FHA+04}.}\\
\multicolumn{3}{p{0.9\textwidth}}{$^2$We use the re-analysed data of \citet{BvLM+10}.}\\
\multicolumn{3}{p{0.9\textwidth}}{$^3$\emph{Spitzer} MIPS; \citet{RYE+04}.}\\
\end{tabular}
\end{table*}
\end{center}

\begin{figure}
\centerline{\includegraphics[width=0.35\textwidth,angle=-90]{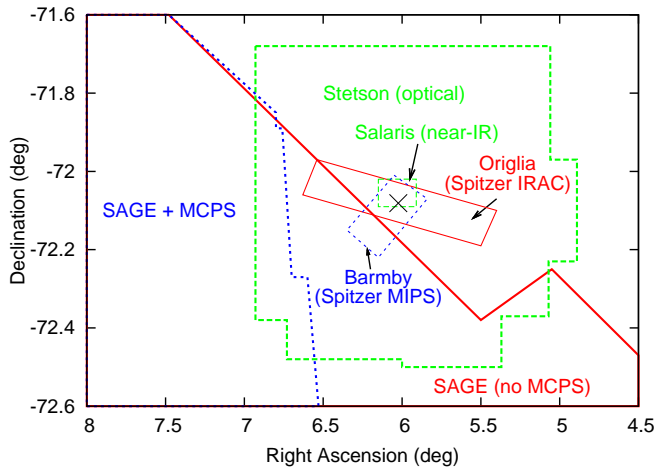}}
\caption{Spatial coverage of the surveys used in this paper. See text for details of each survey. The cross marks the cluster centre.}
\label{CoverageFig}
\end{figure}

As 47 Tuc covers a large area on the sky (its tidal radius is 43$^\prime$; \citealt{Harris96}), a uniform set of photometric data is not available for every star in the cluster. The photometric data used in the SEDs come from a variety of different sources covering different fields of view. These data are summarised in Figure \ref{CoverageFig} and Table \ref{CoverageTable}. They include data taken with the \emph{Spitzer Space Telescope's} \citep{WRL+04} four Infrared Array Camera (IRAC; \citealt{FHA+04}) bands, at 3.6, 4.5, 5.8 and 8 $\mu$m from the Surveying the Agents of Galaxy Evolution in the Small Magellanic Cloud (SAGE-SMC) program \citep{GMB+09}. At the distance of 47 Tuc, the 1.66--1.98$^{\prime\prime}$ full-width at half-maximum (FWHM) of the point spread function (PSF) of IRAC provides a resolution of 0.036--0.043 pc, with a dynamic range sufficient to detect the cluster's HB stars. Further IRAC data of the cluster core were presented in \citet{ORFF+07} (\emph{Spitzer} PID20298; PI: R.~Rood). Here, we use the re-analysed data from \citet{BvLM+10}. For this second IRAC dataset, we used the ``deep'' photometry when [3.6] $>$ 11 mag, and the ``shallow'' photometry otherwise.

Longer-wavelength 24-$\mu$m data was sourced from the Multiband Imaging Photometer for \emph{Spitzer} (MIPS; \citealt{RYE+04}) data from \citet{BBW+09}, as the SAGE-SMC MIPS data do not cover the cluster. 

Near-IR $JHK_{s}$-band photometry for the entire cluster was taken from the 2-$\mu$m All Sky Survey (2MASS; \citealt{SCS+06}). The 2MASS image near the cluster core is heavily blended, and the recorded photometric fluxes are artificially raised by blending of both partially-resolved and unresolved stars. To solve this issue, we preferentially used the $JHK_{s}$-band photometry from \citet{SHO+07} over 2MASS. These data cover a $4{\farcm}9 \times 4{\farcm}9$ region around the cluster core. These data are of substantially better resolution ($<0{\farcs}9$) and consequently suffer less from source confusion and blending. The ``deep'' Salaris photometry was used in preference to the ``shallow'' when $K_{s} > 12.5$ mag. Some stars, in the southern half of the cluster core, were not covered by the Salaris data: many of these therefore have abnormally high 2MASS fluxes, and therefore scatter to higher luminosities. Here they can be confused with AGB stars, though they typically lie slightly above the AGB branch (their locus lies on the dividing line between regions (1) and (2) in the HRD shown later).

The mid-IR \emph{Spitzer} photometry is complemented by optical data from the Magellanic Clouds Photometric Survey (MCPS; \citealt{ZHT+02}), which contains Johnson $UBV$-band and Gunn $i$-band photometry of the region. Once again, this only covers the south-east of the cluster. To cover the remainder of the cluster, we used the Johnson $UBV$-band and Cousins $I_C$-band photometry of \citet{SHO+07}, and the Johnson $UBVI_J$-band\footnote{We refer to the Johnson $I$ band as $I_J$ throughout to avoid confusion with the other $I$-band data in this work.} photometry of \citet{Stetson00}.

\subsection{Creation of a master catalogue}

These data were combined using {\sc daomatch/daomaster} \citep{Stetson93} and objects detected in only one filter (assumed to be bad data) were dropped. The final source list contains photometry for 104\,153 stars located within 30$^\prime$ of the cluster core (00$^h$24$^m$05.2$^s$ --72$^\circ$04$^\prime$51$^{\prime\prime}$ --- \citealt{Harris96}).

From this source list, a subset of sources were selected which had enough photometry to determine their temperatures and luminosities by fitting their SEDs. We selected stars with at least four photometric bands with flux measurements, the bluest of which must be in the optical ($I$-band or shorter wavelength) and the reddest of which must be in the IR ($K_{s}$-band or longer wavelength). This reduced the source list to 47\,727 stars with usable photometry.

\subsection{The data reduction process}
\label{ReductionSect}


\begin{figure*}
\centerline{\includegraphics[width=0.61\textwidth,angle=-90]{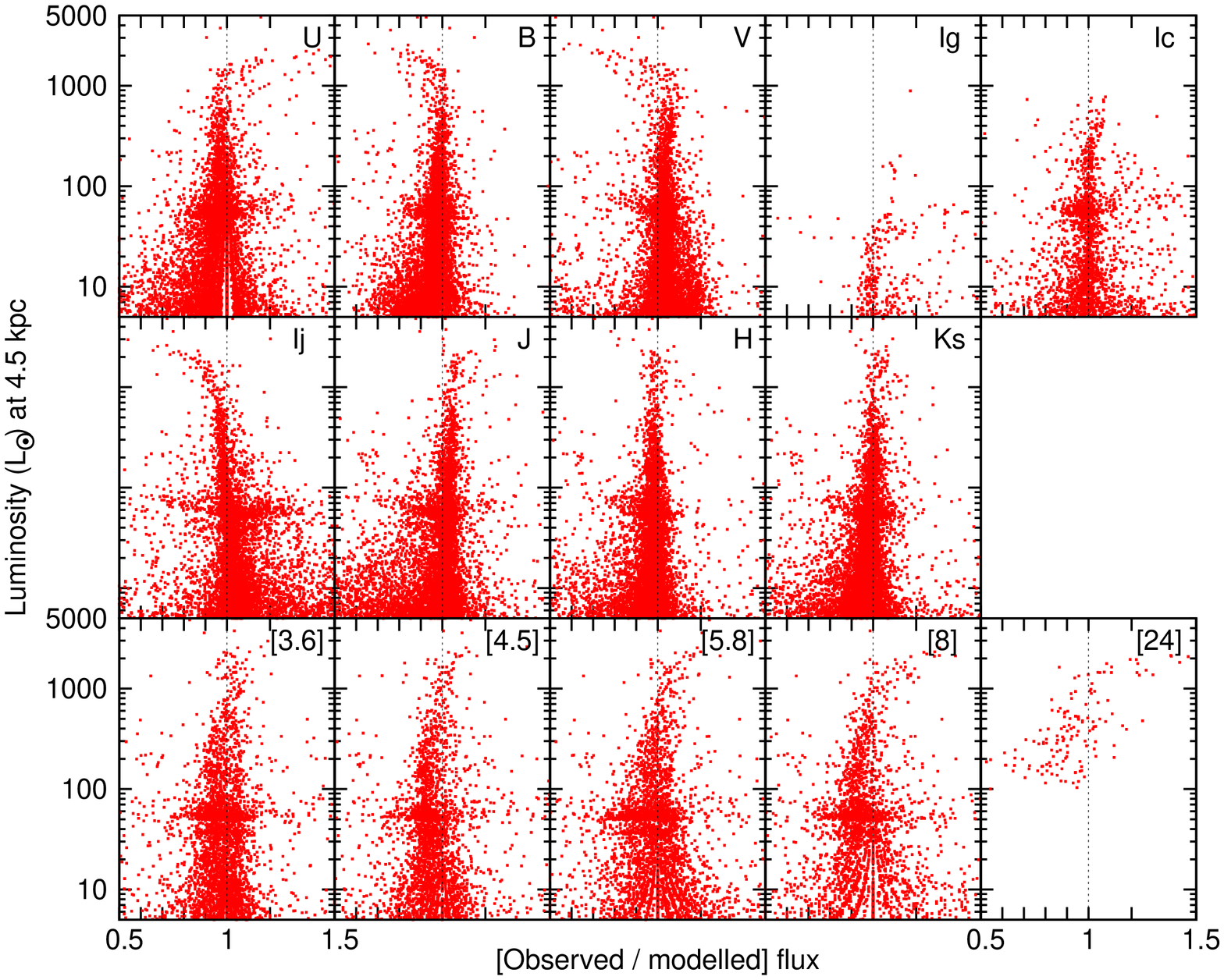}}
\centerline{\includegraphics[width=0.61\textwidth,angle=-90]{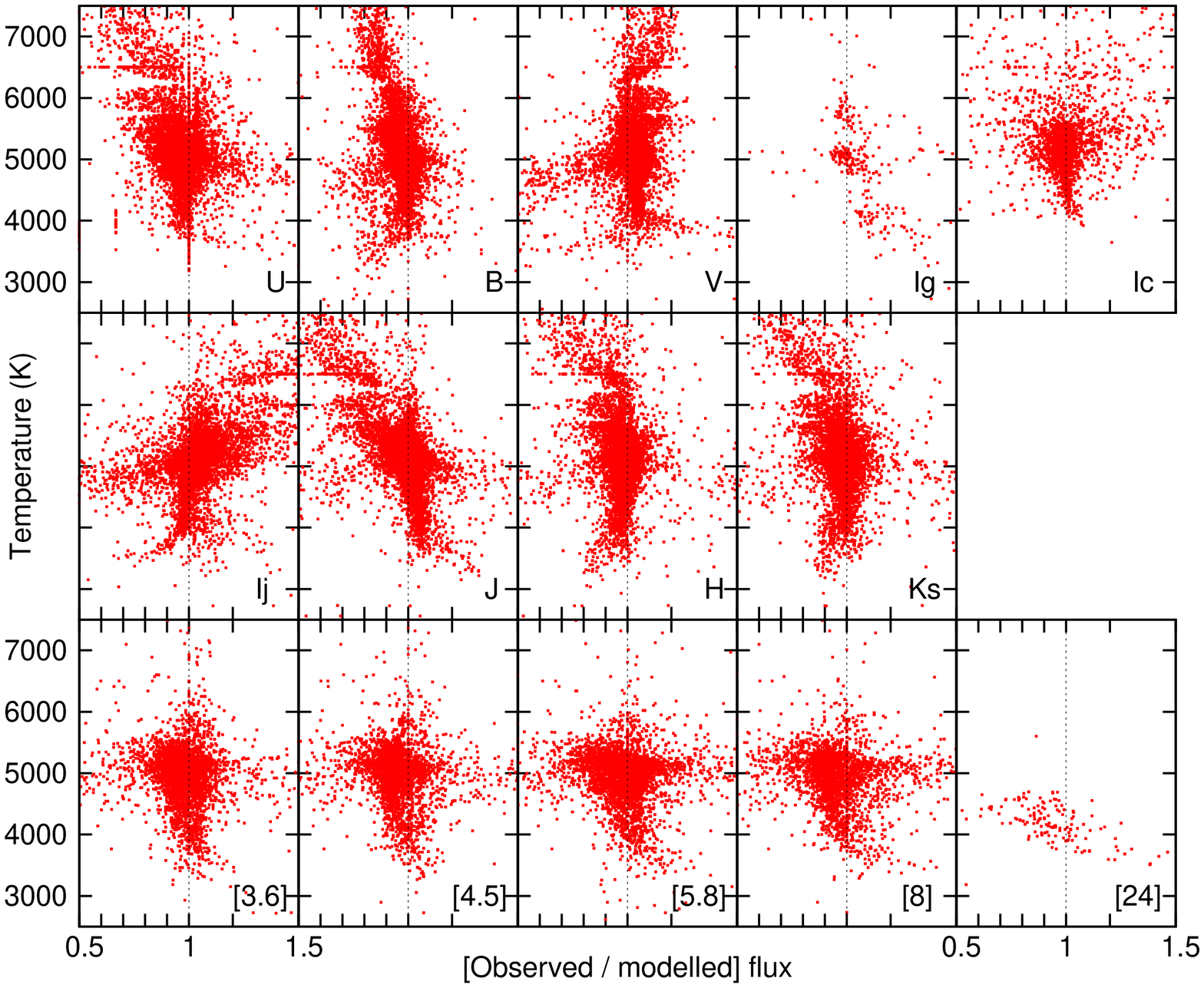}}
\caption{Differences between observed and SED-fitted model fluxes as a function of luminosity (top panels) and temperature (bottom panels, where only stars with $L > 5$ L$_\odot$ are shown).}
\label{OEFig}
\end{figure*}

\begin{center}
\begin{table}
\caption{Adopted central wavelengths and zero points for the photometric filters used.}
\label{ZeroPointTable}
\begin{tabular}{lcc}
    \hline \hline
Filter	& Central wavelength	& Zero point	\\
\ 	& (nm)			& (Jy)		\\
    \hline
$U$	& 350	& 1659	\\
$B$	& 442	& 4130	\\
$V$	& 550	& 3810	\\
$i$	& 786	& 2427	\\
$I_C$	& 810	& 2520	\\
$I_J$	& 880	& 2635	\\
$J$	& 1240	& 1602	\\
$H$	& 1650	& 1010	\\
$K_{s}$	& 2160	& 630	\\
$[3.6]$	& 3600	& 280.9	\\
$[4.5]$	& 4500	& 179.7	\\
$[5.8]$	& 5800	& 115.0	\\
$[8]$	& 8000	& 64.13	\\
$[24]$	& 24000	& 7.14	\\
    \hline
\end{tabular}
\end{table}
\end{center}

SEDs were created following the method used in Papers I and II. Briefly, this involves comparing broadband photometry to artificially-dereddened model spectra which have been convolved to the photometric filters of the observations. Here, we again use the {\sc marcs} models of Paper I (see also \citealt{GBEN75,GEE+08}). Interpolating between these models determines the stellar temperature, and the multiplication of the model flux required to match the observed flux determines the stellar luminosity. We \emph{initially} assume that the distance and reddening to 47 Tuc are 4500 pc and $E(B-V) = 0.04$ mag, respectively. We later show by isochrone fitting that these values are indeed appropriate (\S\ref{SectIsos}).

We have revised the process used in Papers I and II to better account for interstellar reddening.  This follows the same procedure as in \citet{MSZ+10}, namely that we use the absorption profiles of \citet{McClure09} to obtain $A_\lambda / A_{\rm Ks}$, assuming that $A_{\rm V} = 3.2 E(B-V)$ and that $A_{\rm V} / A_{\rm Ks} = 7.75$. While the change in process does not significantly affect the stellar parameters we derive, it does better account for the 10-$\mu$m interstellar absorption peak. This can affect the mass-loss rates we derive, though is unlikely to be significant for the small extinction toward the cluster.

Obtaining the correct filter transmissions is important when obtaining accurate stellar parameters. Differences can occur in the transmission efficiency of the filter itself, the CCD, the telescope and (particularly) the atmosphere. We can test whether our filter transmissions and stellar atmosphere models are correct by plotting the ratio of the observed flux to the flux of our stellar model for each star (Figure \ref{OEFig}), which is very sensitive to errors. Incorrect zero points for a particular magnitude system manifest themselves as global offsets from unity. Incorrect filter transmissions lead to a very strong temperature dependance in the above ratio. We list the adopted zero points in Table \ref{ZeroPointTable}, which gives the approximate corresponding wavelengths for each filter for reference.

Errors in one band will propagate themselves in reverse onto neighbouring bands, making identifying sources of error difficult. Such issues exist for the very hottest and coolest (most-luminous) stars ($>$6500 K and $<$3900 K, respectively). At low temperatures, dynamic atmospheric processes and incomplete molecular opacities in our model atmospheres introduce variations from the model. Conversely, at high temperatures, our models are limited to 6500 K and do not cover the very hottest stars. Stars offset in Figure \ref{OEFig} in all bands may also be Galactic foreground or background SMC field stars. These are not well modelled as their metallicities and surface gravities are substantially different from those of the cluster.

Several other systematics are visible in Figure \ref{OEFig}. These include:
\begin{list}{\labelitemi}{\leftmargin=1em}
\item Quantisation errors among the lowest-luminosity stars, which manifest themselves as fan-shaped spreads visible in the $U$ and 8-$\mu$m filters, but are negligible among the hottest stars.
\item Poor convergence of models around 6250 K. This only affects the hotter stars which have no mid-IR photometry, for which we cannot therefore determine mid-IR excess.
\item Increased scatter in the HB stars around 50 L$_\odot$. This is due to unknown processes (though could indicate an abundance spread) and does not affect any giant branch dust producers.
\item Small ($\sim$1\%) remaining systematic under- or over-estimates of flux in several filters. The colour-based detection criterion we use (\S\ref{SectMdot}) are chosen to maximise detection of IR excess in such cases, but the amplitude of these effects is much smaller than the $>$0.1 mag ($\gtrsim$10\%) excesses we are looking for.
\end{list}
We therefore do not expect these systematics to affect our detection of circumstellar dust.

Problems caused by unresolved blending in the cluster core are visible in several bands. As noted in the previous section, this is a particular concern in the 2MASS data, where there is considerably more scatter in the fitted parameters and observed:expected flux ratios.

Table \ref{ParamTable} contains the list of stellar parameters determined for the stars.

\begin{center}
\begin{table}
\caption{Stellar parameters as determined by SED fitting. A full electronic table is available online.}
\label{ParamTable}
\begin{tabular}{rcccc}
    \hline \hline
Sequence & RA (deg)	& Dec (deg)	& Temperature	& Luminosity \\
Number   & (J2000)	& (J2000)	& (K)		& (L$_\odot$) \\
    \hline
1 & 4.641571 & --72.328035 & 6147 & 258.4 \\
2 & 4.685675 & --72.365643 & 4449 & 56.53 \\
3 & 4.706252 & --72.387843 & 5399 & 1.625 \\
4 & 4.722603 & --72.374671 & 4500 & 1.228 \\
5 & 4.725042 & --72.366471 & 5308 & 1.057 \\
\nodata	& \nodata	& \nodata	& \nodata	& \nodata \\
    \hline
\end{tabular}
\end{table}
\end{center}

\subsection{Accuracy of the fits}
\label{AccSect}

\subsubsection{Analysis of departures from the models}

\begin{figure}
\centerline{\includegraphics[width=0.61\textwidth,angle=-90]{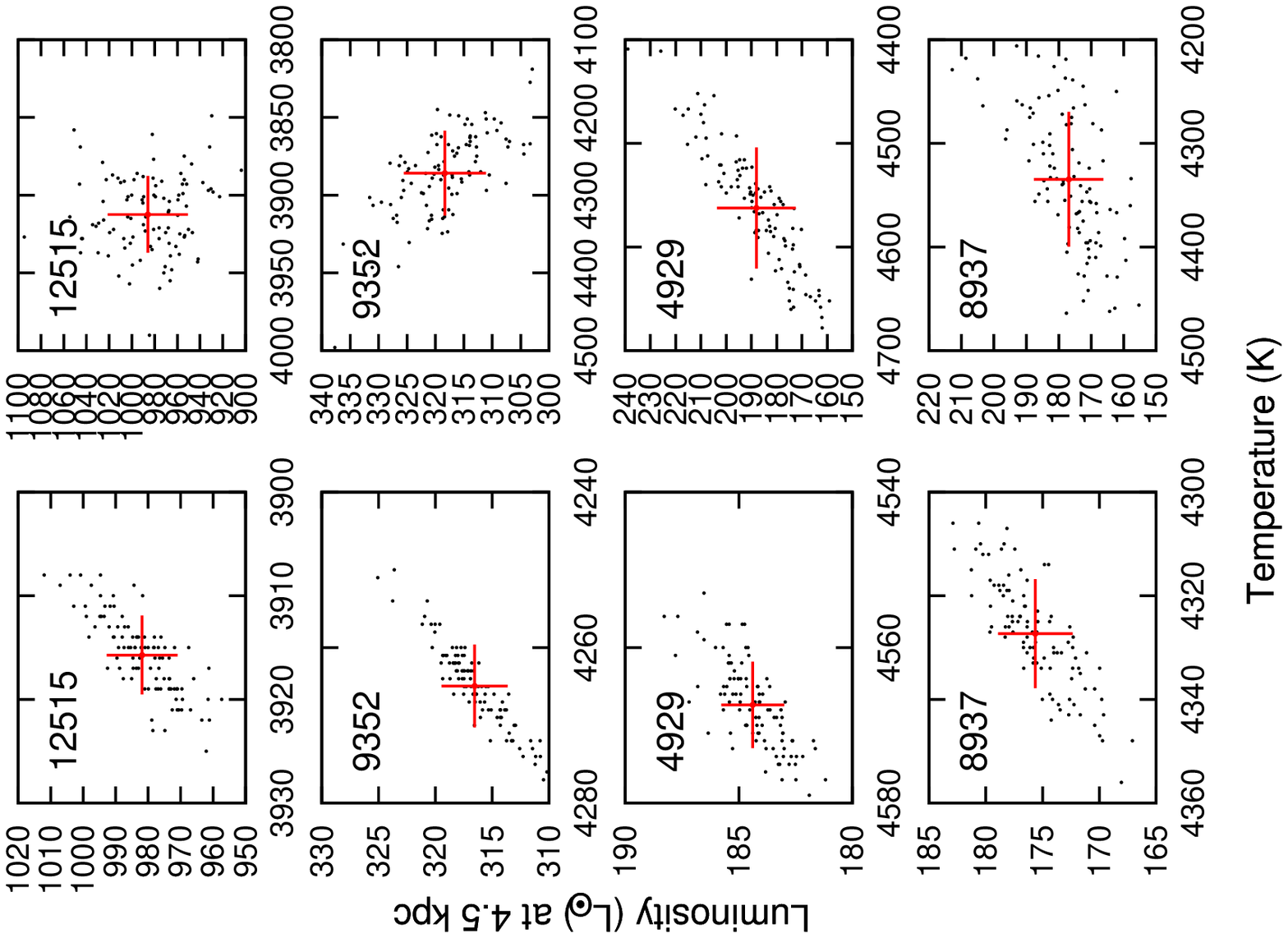}}
\centerline{\includegraphics[width=0.61\textwidth,angle=-90]{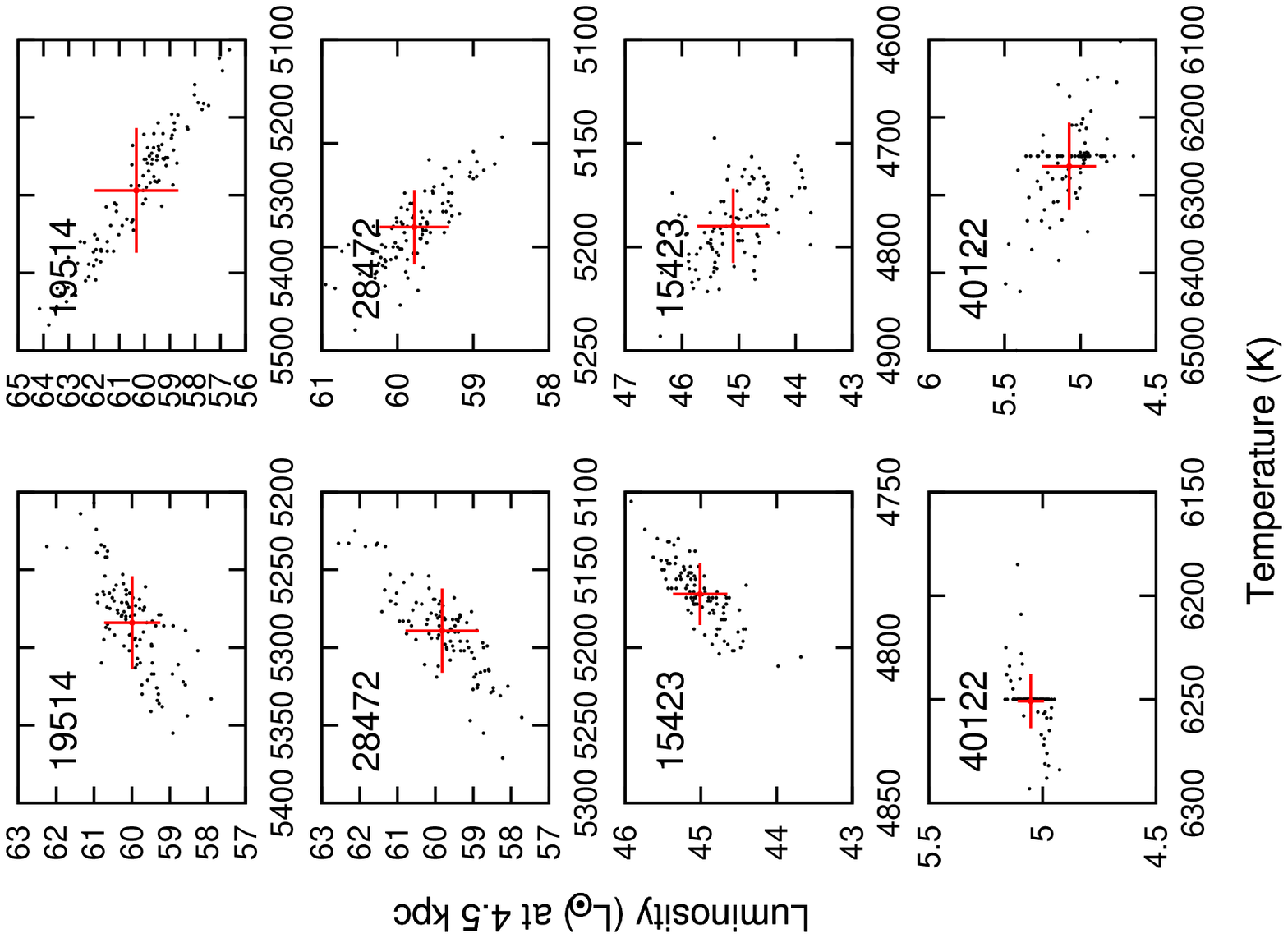}}
\caption{Monte-Carlo fitting results, showing (left) photometric errors only and (right) errors derived from departures from the model. Red bars show the 1-D standard deviations of the distributions. Labels denote sequence numbers of the objects (Table \ref{ParamTable}), chosen to be representative of their respective parts of the HRD.}
\label{MCFig}
\end{figure}

\begin{center}
\begin{table*}[t]
\caption{Stars chosen for Monte-Carlo simulations.}
\label{MCTable}
\begin{tabular}{rccr@{}l@{}lr@{}l@{}lcl}
    \hline \hline
Sequence & RA (deg)	& Dec	(deg)	& \multicolumn{3}{c}{Temperature$^1$}	& \multicolumn{3}{c}{Luminosity$^1$} & Coverage	& Notes \\
Number   & (J2000)	& (J2000)	& \multicolumn{3}{c}{(K)}		& \multicolumn{3}{c}{(L$_\odot$)}  & \ 	& \ \\
    \hline
12515 & 5.9604583 & --72.0725000 &   3916 &$\pm$4  &$\pm$25 &   982   &$\pm$11  &$\pm$35   &   $B-[24]$  & Upper RGB/AGB \\
9352  & 5.9282500 & --72.1920278 &   4265 &$\pm$5  &$\pm$55 &   316   &$\pm$3   &$\pm$7.2  &   $B-K_{s}$   & Central RGB/AGB \\
4929  & 5.7577500 & --72.1565278 &   4567 &$\pm$6  &$\pm$59 &   184   &$\pm$1.4 &$\pm$16   &   $B-[8]$   & Lower AGB \\
8937  & 5.9237083 & --72.0840556 &   4329 &$\pm$10 &$\pm$65 &   175   &$\pm$3   &$\pm$11   &   $B-[24]$  & RGB \\
19514 & 6.0277917 & --72.0754167 &   5283 &$\pm$30 &$\pm$80 &    60.0 &$\pm$0.7 &$\pm$1.7  &   $B-K_{s}$   & HB (core) \\
28472 & 6.1138333 & --72.0774722 &   5191 &$\pm$27 &$\pm$18 &    59.8 &$\pm$1.0 &$\pm$0.5  &   $B-K_{s}$   & HB (not core)\\
15423 & 5.9881250 & --71.9649722 &   4782 &$\pm$10 &$\pm$36 &    45.0 &$\pm$0.4 &$\pm$0.6  &   $B-K_{s}$   & RGB clump \\
40122 & 6.5137500 & --72.2503611 &   6250 &$\pm$13 &$\pm$56 &     5.05&$\pm$0.06&$\pm$0.18 &   $B-[4.5]$ & MSTO \\
    \hline
\multicolumn{11}{p{0.9\textwidth}}{$^1$First error encompasses photometric errors, second error encompasses modelling errors.}
\end{tabular}
\end{table*}
\end{center}

As noted in \S\ref{ReductionSect}, the goodness-of-fit of our SED models depends strongly on the zero points and filter transmissions used. While Figure \ref{OEFig} shows that the models agree well with the observations between temperatures of 3900 and 6500 K, stars lying outside of this range are subject to some error. To assess the magnitude of this error, a representative selection of stars was taken and the data from the most-affected filters ($U$, $B$ and $I_J$) were removed. This had a negligible ($<$5 K) effect on the stellar temperatures of the stars in the temperature range 3900--6500 K, with temperatures changing slightly more for stars outside this range. The coolest stars have temperature uncertainties caused by poorly-matching models of order tens of degrees, but these stars are also strongly variable. Any `instantaneous' measure of temperature such as this will therefore not necessarily be representative of the mean stellar temperature to this degree of accuracy.

A formal estimate of the error may be calculated using `simulated' observations using the Monte-Carlo method, whereby the flux in each band is set to be the observed value plus-or-minus an error sampled from a Gaussian probability distribution with $\sigma$ identical to the reported photometric error. This suffers two problems: (1) it would take prohibitively long to perform this test for every star and (2) it only takes into account the photometric error, not the modelling error. We have therefore selected a small sample of stars from across the HRD, and performed Monte-Carlo simulations to determine the errors in their parameters. We do this twice: the first accounts for only the photometric error, as described above; the second replaces the Gaussian $\sigma$ with the deviation of the observations from the model.

The stars with which we perform this test are listed in Table \ref{MCTable} and the results are shown in Figure \ref{MCFig}. It is firstly clear that the temperature and luminosity we derive are interdependent: spread along the RGB track appears determined primarily by errors in the optical data, while spread perpendicular to the RGB track appears determined primarily by errors in the IR data. It is also clear that the photometric errors are considerably less than the total error as estimated by differences between the observations and stellar atmosphere models. As we note in \S\ref{XSStarsSect}, this is probably due to under-reported errors due to image artifacts and blending from sources within a few arcseconds. In the right-hand panels of Figure \ref{MCFig}, we note that those stars with no data longward of $K_{s}$-band have significantly more scatter perpendicular to the giant branch: \emph{Spitzer} data are therefore crucial in determining accurate stellar parameters, even for warm stars.

Varying a star's fundamental parameters can also have an effect on the values we achieve. Likely variations for \emph{bona fide} 47 Tuc cluster members are somewhat smaller than the corresponding errors in temperature or luminosity caused by inaccuracies in the model, which are of order 1--2\% for temperature and 2--9\% in luminosity. By varying [Fe/H] between solar and --1.4 we produced differences of similar magnitude, varying the mass from 0.6--1.0 M$_\odot$ produced differences of order 0.3\% in luminosity, and varying $E(B-V)$ between 0 and 0.12 mag produced luminosity changes of $<$5\%. The largest uncertainty, at least in terms of luminosity, appears to be in the distance to the cluster.

\subsubsection{Comparison with other works}
\label{CompSect}

In order to determine the accuracy of our SED fits, we compare our temperatures with a number of spectroscopically-derived temperatures. We do not compare the luminosities, as the fitting procedure means that the luminosity is directly determined from the model after the temperature fit has been made. The resulting discrepancies are listed below:
\begin{list}{\labelitemi}{\leftmargin=1em}
\item \citet{CGB+04} --- nine early-RGB stars ($\approx$5100 K), temperatures agree within errors: on average they are 21 K (0.4\%) warmer in our SED fits than their spectroscopic temperatures, cf.\ 108 K for the standard deviation of the temperature differences.
\item \citet{MvL07} --- nine upper RGB/AGB stars ($<$4000 K, $>$1100 L$_\odot$), temperatures also agree within errors, averaging 47 K (1.3\%) cooler in our SED fits (st.\ dev.\ 311 K).
\item \citet{KM08} --- eight central RGB stars (4200--4500 K), individual temperatures differ at the 2$\sigma$ level. As a whole, the stars average 25 K (0.6\%) warmer in our SED fits, smaller than the standard deviation of the differences between the two papers (31 K).
\item \citet{WCMvL10} --- five of the spectra used in \citet{MvL07}, individual temperatures differ at the 4$\sigma$ level, averaging 231 K (6.1\%) cooler in our SED fits (st.\ dev.\ 108 K).
\end{list}

The exceptionally low standard deviation of the \citet{KM08} temperatures provides a good check of the accuracy of both our measurements and theirs. It is worth noting that the discrepancy between our results and those of \citet{CGB+04} lies in the accuracy of our photometry, not their spectroscopic determinations. A systematic offset between our SED fitting and conventional spectroscopic techniques is likely limited to $<$30 K ($<$0.7\%) in temperature and $<$2.8\% in luminosity.


\subsection{The Hertzsprung--Russell diagram}
\label{HRSect}

\begin{figure*}
\centerline{\includegraphics[width=0.6\textwidth,angle=-90]{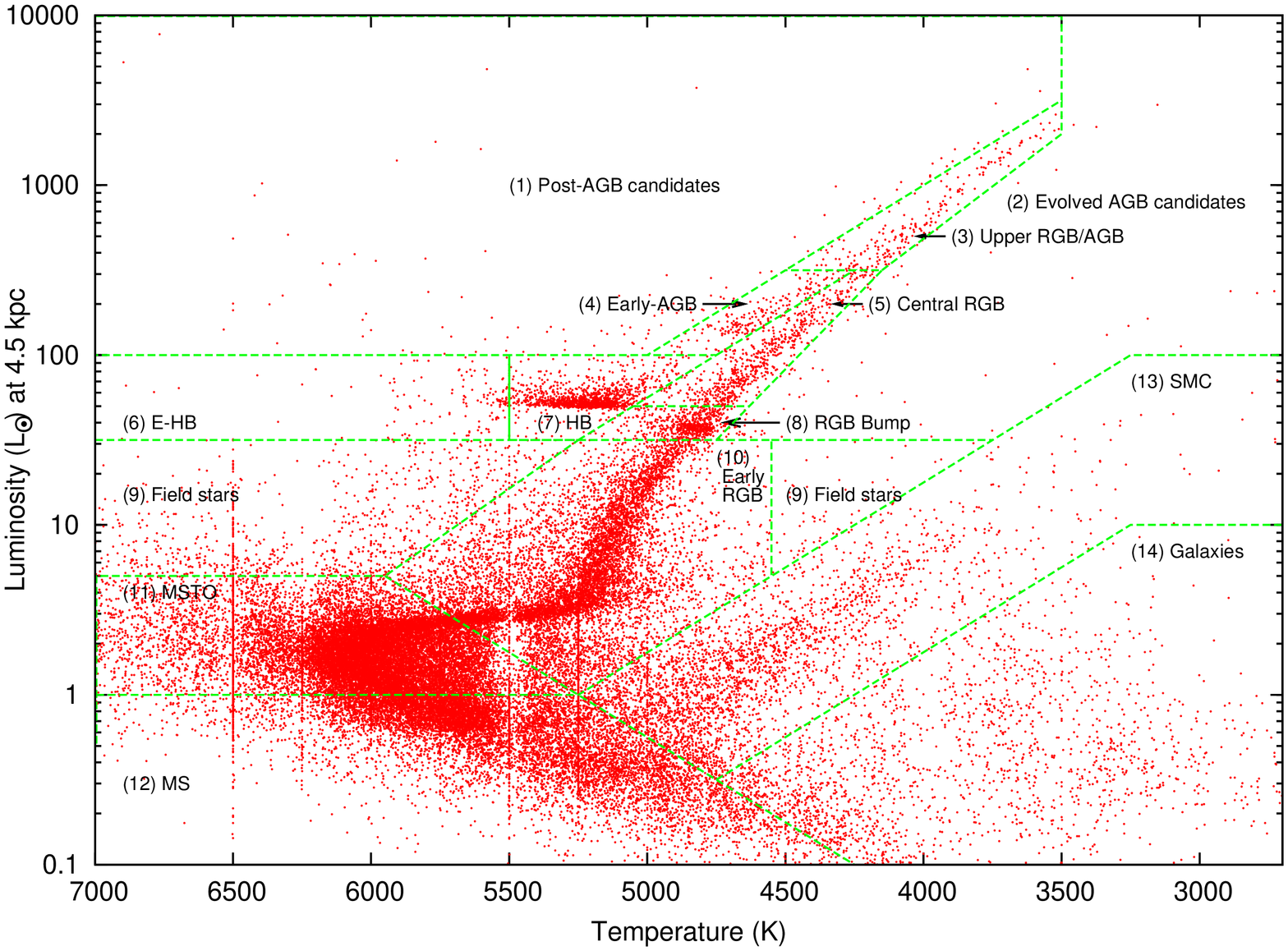}}
\centerline{\includegraphics[width=0.6\textwidth,angle=-90]{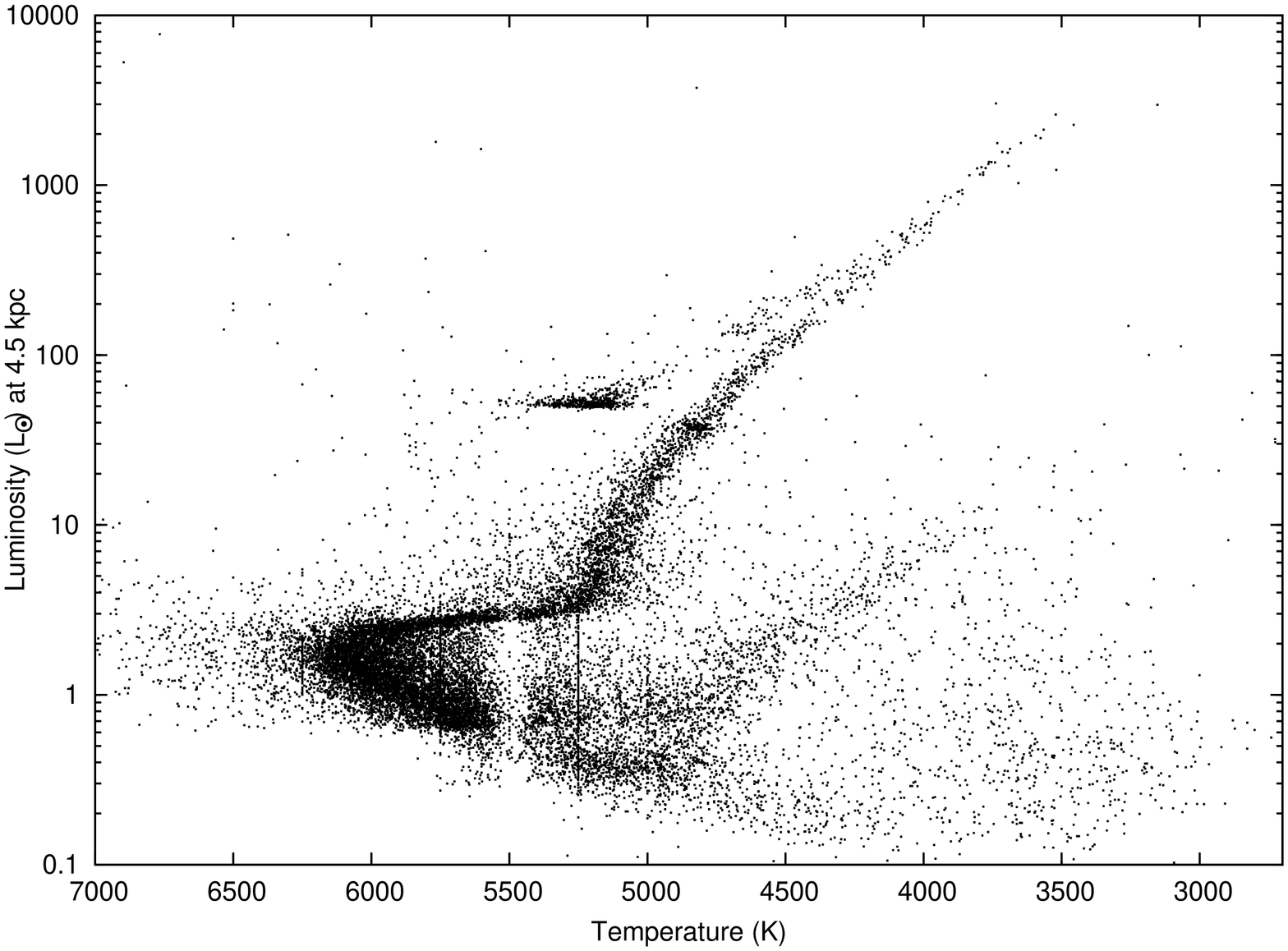}}
\caption{HRD of stars in the direction of 47 Tuc, showing evolutionary and spatial components. The top panel contains all stars, the bottom panel contains only stars more than 2$^\prime$ from the cluster core, which have photometric data in the optical, near-IR and mid-IR. Abbreviations are discussed in the text.}
\label{HRDFig}
\end{figure*}

Having determined the stellar parameters for the above objects, the data were then compiled into an HRD, presented in Figure \ref{HRDFig}.

The HRD shows a number of features, mostly identifying our target stars as post-main-sequence stars in 47 Tuc. Also visible on the HRD is a background sequence of stars belonging to the SMC (13). This merges with the main sequence (MS) of 47 Tuc at roughly the luminosity where the SMC's HB and red clump would be. Regions (9) and (14) contain foreground and background objects of indeterminate distance. Typically, we would expect region (9) to contain stars in the Galactic disc (47 Tuc lies at Galactic co-ordinates $l = 305.90^{\circ}$, $b = -44.89^{\circ}$ --- \citealt{Harris96}) and region (14) to contain background galaxies (\citealt{BMvL+08}; Paper I). In this particular sample, however, region (14) may also contain some SMC young stellar objects (YSOs) and cool main sequence objects. We do not expect many YSOs, however, as our region of interest lies away from the bulk of the galaxy's star-forming regions. The main sequence stars at this temperature are mostly below the sensitivity threshold of our data.

Vertical artifacts are also visible on the diagram. These occur every 250 K and correspond to the temperature grid of our model spectra. They are generally caused by poor-quality optical photometry in low-luminosity stars. Especially important are discrepancies between the observed $U$- and $B$-band photometry and those predicted by the {\sc marcs} model spectra, hence the increased prevalence of artifacts at higher temperatures where more flux is emitted in $U$ and $B$. As our models only extend to 6500 K, we must assume that stars with temperatures derived to be $\gtrsim$6500 K have progressively more-uncertain parameters. We have therefore limited our study to stars with $T \leq 7000$ K, of which there are 46\,398.

Several important evolutionary points (numbered in Figure \ref{HRDFig}) can be noted at this point, which we use in the next section to fit stellar isochrone models. These are: the main-sequence turnoff (MSTO; 11 in Figure \ref{HRDFig}), the start of the RGB (10), the RGB bump (8), the upper RGB (5), the RGB tip (top of (2)), the base of the HB (bottom of (7)) and the start of the AGB (bottom of (4)). A review of these evolutionary processes, with particular emphasis on globular clusters, can be found in \citet{McDonald09}.

Circumstellar dust reprocesses optical light into the IR, causing an apparent cooling of the star's effective temperature. This moves the star to the right on the HRD, moving into region (2). Throughout their time on the RGB, stars lose mass through stellar winds. For the majority of the time, these appear to be dustless winds driven from the chromosphere (\citealt{MvL07}; Paper I), but if the IR excesses reported by \citet{ORFF+07} are true, we may see some low-luminosity stars in this region by virtue of their circumstellar dust. This issue is confused, however, by the presence of field stars, and of stars with poorly-determined photometry.

Our HRD contains a well-denoted and well-separated early-AGB. By its very nature, however, the AGB asymptotically approaches the RGB, and the two become indistinguishable above $\sim$300 L$_\odot$. More-rapid evolution on the AGB means that there are 4--5 times more stars per unit luminosity on the RGB than AGB. 

Helium fusion on the upper AGB also becomes increasingly volatile: its violence can increase a star's luminosity dramatically (up to a factor of several) for a short period of time ($10^2 \sim 10^3$ years), termed a thermal pulse (TP; e.g.\ \citealt{Zijlstra95,WWSS08}). Such stars may scatter above the main giant branch in the HRD, into region (1), but this area is confused by normal AGB stars with poor photometry.

Stars on the TP-AGB are also unstable to pulsation via the $\kappa$-mechanism \citep{Ulmschneider98}, where harmonic oscillations are stochastically excited. This causes semi-regular variability at first, then more-regular and increasingly-violent pulsation until the AGB tip, where the stellar atmosphere disperses. This variability will cause some horizontal and vertical scatter of these stars in the HRD, as we typically only use one epoch of photometry for every filter.


\section{Isochrone fitting}
\label{SectIsos}

\subsection{Introduction}
\label{SectIsosIntro}

By fitting stellar isochrones to our data, we can re-determine the basic parameters for the cluster as a whole, namely its distance, reddening and age. To do this, we must make assumptions about the cluster's metallicity, helium fraction and $\alpha$-element enhancement. The accuracy of this determination depends both on the systematic errors present in our data (\S\ref{AccSect}), and the validity of the assumptions we make. To minimise systematic errors, we have used the `cleaned' subset of stars presented in the bottom panel of Figure \ref{HRDFig}, to the exclusion of stars with poor photometry and those in the immediate cluster core.

In this study, we compare our HRD to two sets of isochrones, both of which were used in Paper I. These are namely the Padova isochrones \citep{MGB+08,BGMN08}\footnote{http://stev.oapd.inaf.it/cgi-bin/cmd} and the Dartmouth isochrones \citep{DCJ+08}\footnote{http://stellar.dartmouth.edu/$\sim$models/webtools.html}.

The CMD interface for the Padova isochrones \citep{MGB+08} does not allow the user to change either the helium content or $\alpha$-enrichment of the isochrones. The YZVAR interface \citep{BGMN08}, however, allows both the helium content and a Reimers-law mass-loss rate \citep{Reimers75} to be varied by the user, though also does not allow the $\alpha$-enrichment to be varied. In this regard, we have simply adopted a fixed metal content, [Z/H], and assumed that metals scale as solar values. The Dartmouth models allow the variation of both helium and $\alpha$-enrichment in fixed steps.

\begin{figure}
\includegraphics[height=0.47\textwidth,angle=-90]{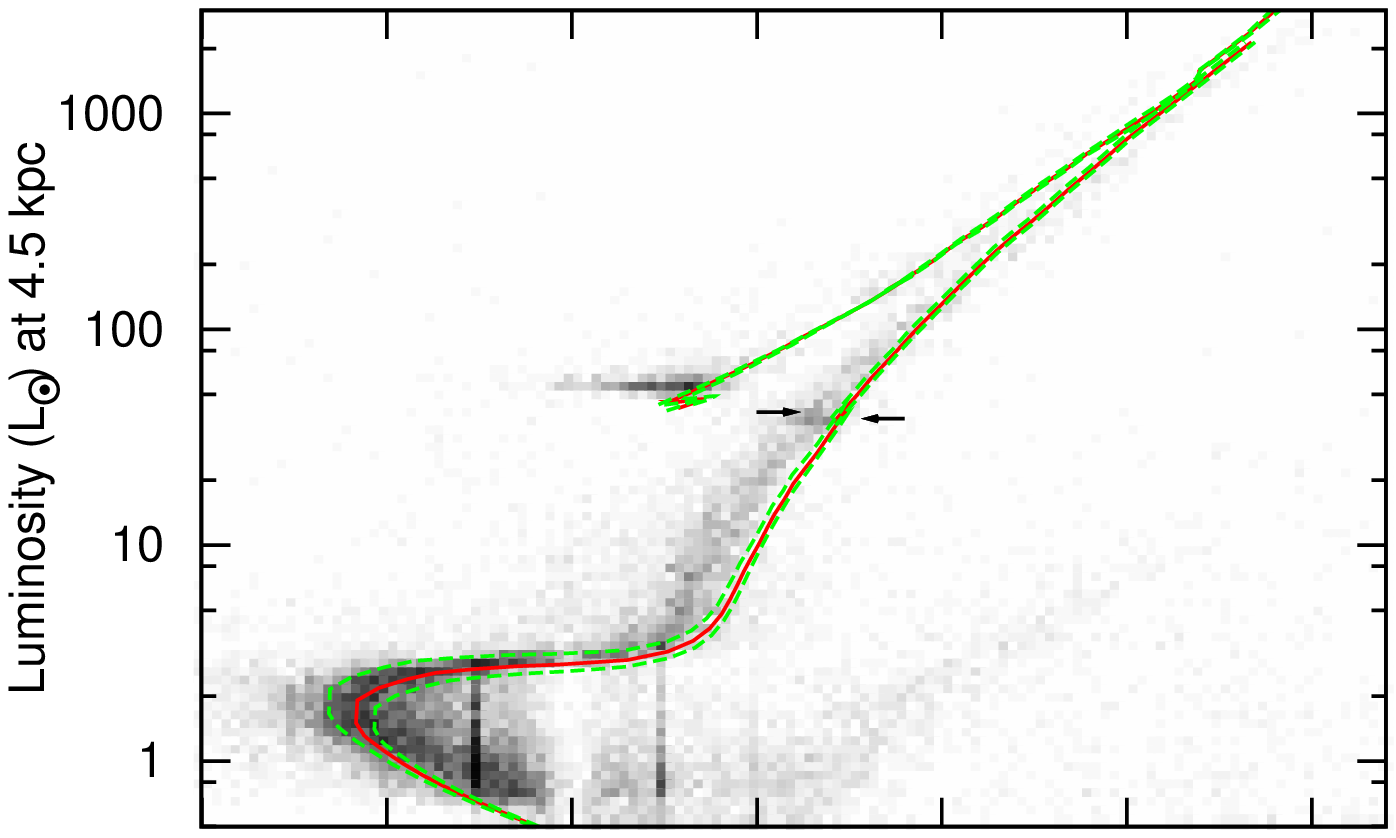}
\includegraphics[height=0.47\textwidth,angle=-90]{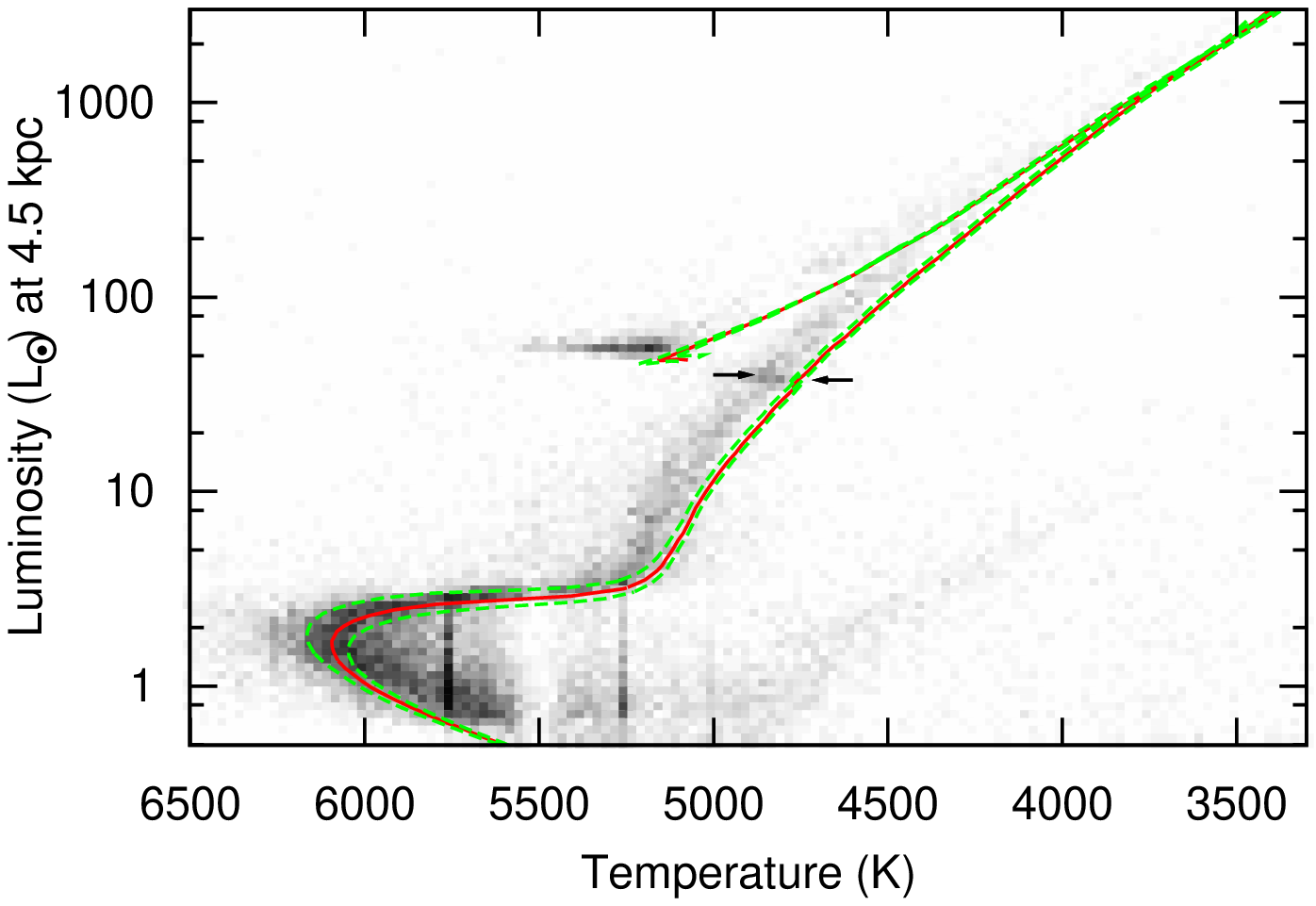}
\caption{Top panel: Padova isochrones generated using the CMD utility, overlaid on a density-mapped HRD. Bottom panel: Padova isochrones generated using the YZVAR utility. Arrows show the locations of the RGB bumps of the isochrones. Details of the isochrones' parameters can be found in \S\ref{SectIsosIntro}.}
\label{IsoPadovaFig}
\end{figure}

\begin{figure}
\includegraphics[height=0.47\textwidth,angle=-90]{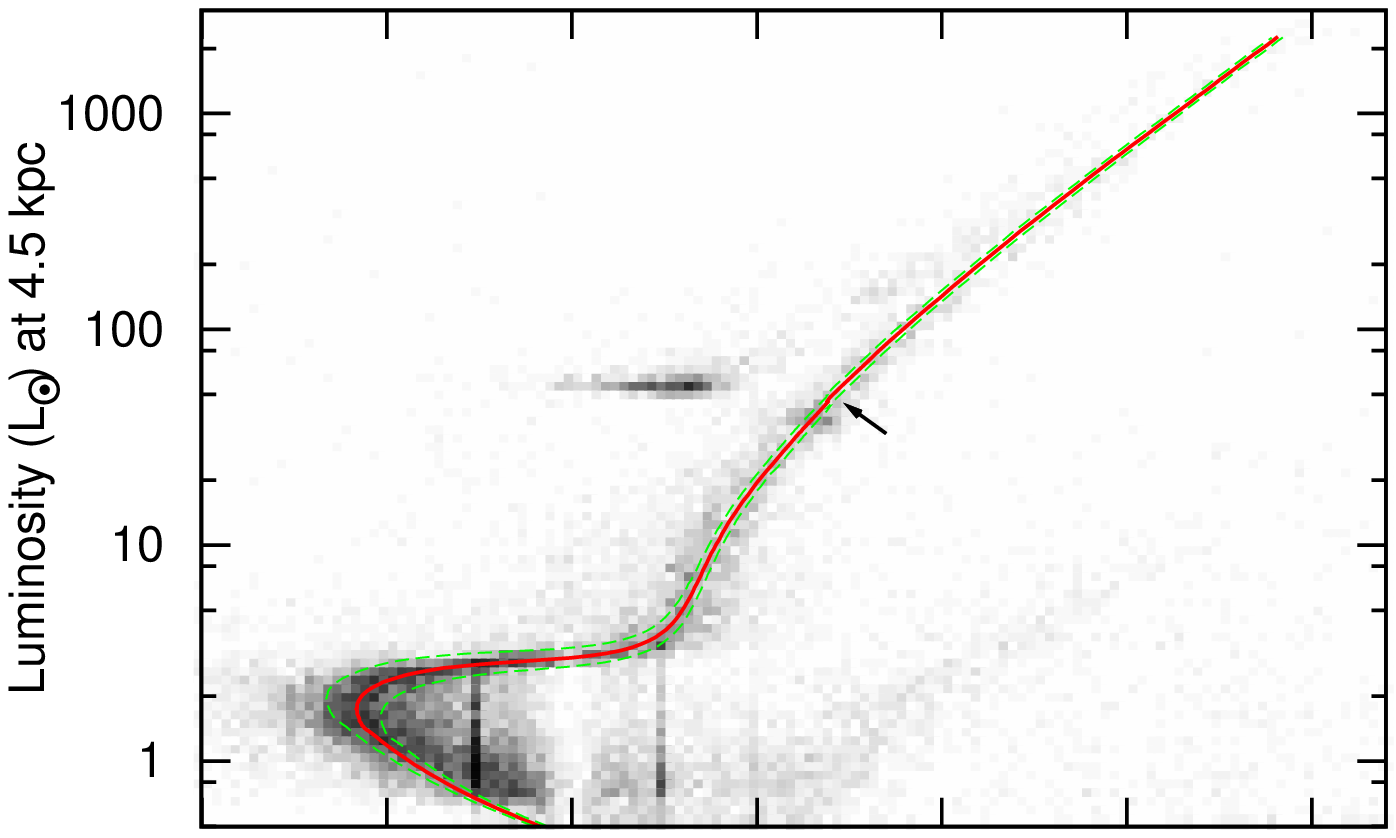}
\includegraphics[height=0.47\textwidth,angle=-90]{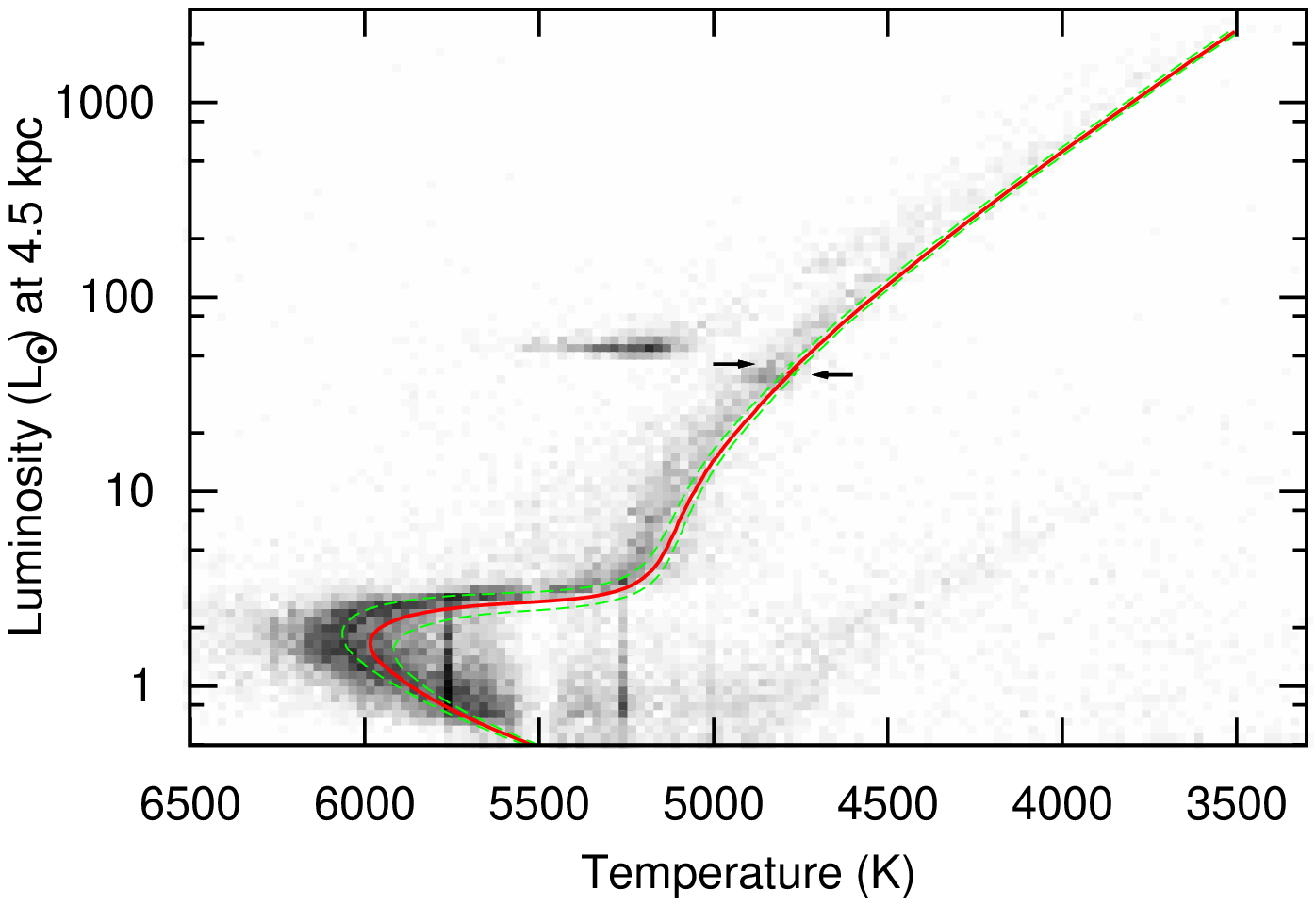}
\caption{As Figure \ref{IsoPadovaFig} for the Dartmouth isochrones. Top panel: isochrones at [$\alpha$/Fe] = 0. Bottom panel: isochrones at [$\alpha$/Fe] = +0.2.}
\label{IsoDartmouthFig}
\end{figure}

Altering the above parameters has the following effects:
\begin{list}{\labelitemi}{\leftmargin=1em}
\item Increasing the helium content of the isochrones steepens the slope of the RGB, though the RGB tip stays in roughly the same place. It also causes an increase in the luminosity difference between the MSTO and the start of the RGB.
\item Increasing the $\alpha$-enrichment accelerates MSTO, meaning this point corresponds to lower luminosities. It also shifts the entire isochrone to cooler temperatures and increases the luminosity of the RGB bump.
\item Increasing the metallicity of the isochrones will shift them to cooler temperatures, as stronger photospheric lines make the atmosphere more opaque.
\item Increasing the fitted distance of the cluster will shift the isochrones toward higher luminosities with respect to the stars.
\item If the cluster's reddening is higher, we will have therefore underestimated the de-reddening correction: further correction then requires the stars to move to higher temperatures and luminosities with respect to the isochrones. This does not affect all stars equally, and hotter stars will require larger corrections than cooler stars, with the result that the HRD becomes horizontally stretched and vertically skewed (see Paper I for the way this is treated).
\item The cluster's age affects the isochrones in a number of ways, with the largest alterations being in the positioning of the MSTO and base of the RGB.
\end{list}

We initially assumed the following parameters:
\begin{list}{\labelitemi}{\leftmargin=1em \itemsep=0pt}
\item Metallicity: [Z/H] = [Fe/H] = --0.7;
\item Helium fraction: $Y$ = 24\% $\approx Y_{\odot}$ \citep{DVL89};
\item $\alpha$-enrichment: [$\alpha$/Fe] $\approx$ +0.2 dex (Dartmouth isochrones only);
\item Distance: $d$ = 4.5 kpc;
\item Reddening: $E(B-V)$ = 0.04 mag; $A_{\rm V} = 0.124$ mag; $R = 3.1$;
\item Age: $t$ = 12.5 $\pm ^{1.2}_{1.4}$ Gyr.
\end{list}
Isochrones using these parameters are shown in Figures \ref{IsoPadovaFig} and \ref{IsoDartmouthFig}. It is clear from these figures that the best-fitting isochrone is that of the Dartmouth models, but at the adopted value of [$\alpha$/Fe] = 0. This provides a remarkably-good fit to the data. The only exception to this is on the upper RGB, above the red clump, where the [$\alpha$/Fe] = +0.2 model fits better (see also later, in Figure \ref{IsoDartmouthHB}). An increase in $\alpha$-element abundances following first dredge-up is not expected, as the stars have not begun helium burning. It could, however, be explained if extra mixing on the RGB occurs around first dredge-up, and if such mixing alters the surface chemistry (e.g.\ \citealt{DV07}; \citealt{KCS10}, and references therein). This effect is shown in more detail in \S\ref{HBMassSect}. Note that the Padova isochrones, which qualitatively fit the upper giant branches and MSTO, are also for [$\alpha$/Fe] = 0.

\subsection{Cluster parameters}

\subsubsection{Metallicity and abundances}

The metallicity and abundance of 47 Tuc's stars is better determined from spectroscopic studies of individual stars than from our HRD (see \S\ref{CompSect} for such studies). As noted, however, the best-fitting isochrone requires [$\alpha$/Fe] = 0. To fit the Padova isochrones by changing the $\alpha$-enhancement (which roughly corresponds to moving the isochrone horizontally in the HRD) requires [$\alpha$/Fe] $<$ 0. For the purposes of isochrone fitting, we have proceeded under the assumption that there is no $\alpha$-element enrichment compared to solar values.

\subsubsection{Reddening}

The de-reddening of the cluster we apply can be altered using the relations listed in Paper I. This gives an approximate solution for different values of $E(B-V)$ without the need to re-analyse the entire dataset. Based again on the Dartmouth [$\alpha$/Fe] isochrones, we find agreement with the accepted value of $E(B-V) \approx 0.04$ mag \citep{Harris96}.

\subsubsection{Age and distance}

\begin{figure}
\includegraphics[height=0.47\textwidth,angle=-90]{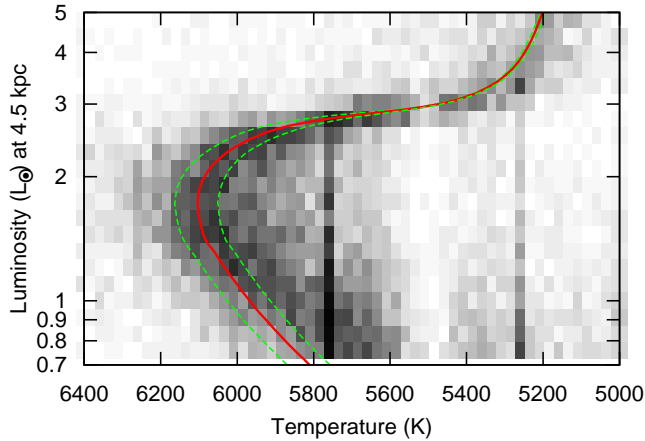}
\caption{As Figure \ref{IsoDartmouthFig}, showing Dartmouth isochrones at [$\alpha$/Fe] = 0, covering the MSTO. Isochrones are placed at 11, 12 and 13 Gyr.}
\label{IsoDartmouthAge}
\end{figure}

Once the metallicity, abundances and reddening of the stars have been set, the age can primarily be determined from the temperature of the MSTO. This is vertically scaled to match the luminosity of the MSTO/RGB base at 5500 K. The parameters of age and distance are thus anti-correlated.

As shown in Figure \ref{IsoDartmouthAge}, isochrones for 12 $\pm$ 1 Gyr fit the data well. An incomplete stellar catalogue of objects below $\sim$1 L$_\odot$ means that the models will fit progressively less well below this luminosity. Isochrones are plotted for distances of 4783, 4611 and 4456 pc for 11, 12 and 13 Gyr, respectively. Adding in the 2.8\% systematic luminosity uncertainty in luminosity (\S\ref{CompSect}), this equates to a distance of 4611 $^{+213}_{-200}$ pc.

Both sets (CMD \& YZVAR) of Padova isochrones yield similar constraints on age. The distances implied for these isochrones are approximately 4\% (180 pc) less than for the Dartmouth isochrones.

\subsubsection{Luminosity function}
\label{LumFnSect}

\begin{figure}
\includegraphics[width=0.35\textwidth,angle=-90]{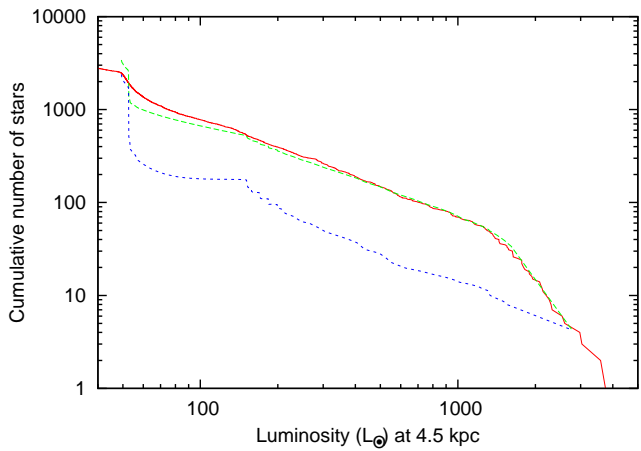}
\caption{The luminosity function for 47 Tuc (solid, red line), along with theoretical Pisa isochrones. Lower, short-dashed line (blue): AGB and HB stars only. Upper, long-dashed line (green): AGB/HB + RGB stars.}
\label{IsoPisaLFnFig}
\end{figure}

Figure \ref{IsoPisaLFnFig} shows the observed and theoretical luminosity functions of 47 Tuc. The observed luminosity function is derived from all stars below 6000 K and 4200 L$_\odot$ (to include the dusty long-period variable star 47 Tuc V1, but exclude known field stars) which do not fall in region 13 (SMC) of the HRD. The theoretical isochrone uses the Pisa stellar evolution models for a 0.90 M$_\odot$ RGB and a 0.65 M$_\odot$ AGB star. The isochrones have been scaled to one star for every 80\,000 years of evolution. It should be noted that the AGB models terminate at 1550 L$_\odot$: we have extrapolated them in Figure \ref{IsoPisaLFnFig} to estimate that there are some 8$\pm$2 AGB stars above the theoretical RGB tip at 2780 L$_\odot$.

Overall, the luminosity function provides a good fit to the data. The departure between 60 and 100 L$_\odot$ is probably due to contamination from field stars and scatter in the HRD from poor photometry. The fact that we do not find as many HB stars at $\approx$50 L$_\odot$ as the models predict suggest that either we do not have a complete sample of all HB stars, or that some of the HB stars have already been ejected from the cluster due to mass segregation (cf. HB lifetime $\approx 190$ Myr; core relaxation time $\approx 91$ Myr; \citealt{Harris96}). This is not readily visible in radial density plots of our sources due to the differing depths of the original surveys we used.

\subsubsection{HB mass via isochrone fitting}
\label{HBMassSect}

\begin{figure}
\includegraphics[width=0.35\textwidth,angle=-90]{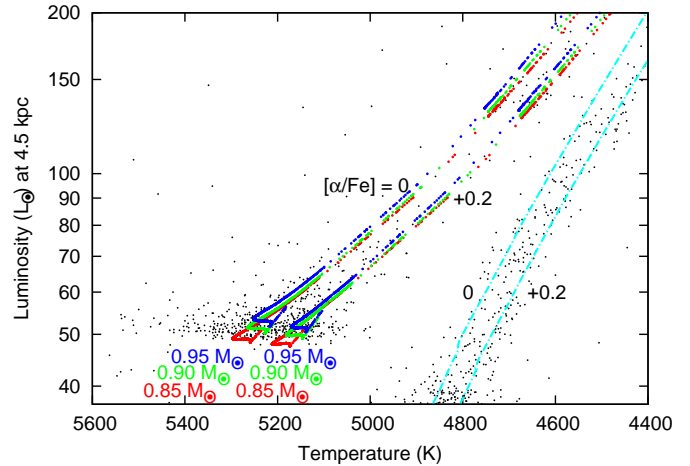}
\caption{The upper RGB, HB and AGB of 47 Tuc (small, black dots). These are overlain with Dartmouth RGB isochrones (blue dashed lines) at [$\alpha$/Fe] = 0 (left) and +0.2 (right), and with similar HB models (large, coloured dots). In order of increasing luminosity, the HB tracks are for stellar masses of 0.85, 0.90 and 0.95 M$_\odot$.}
\label{IsoDartmouthHB}
\end{figure}

\begin{figure}
\includegraphics[width=0.35\textwidth,angle=-90]{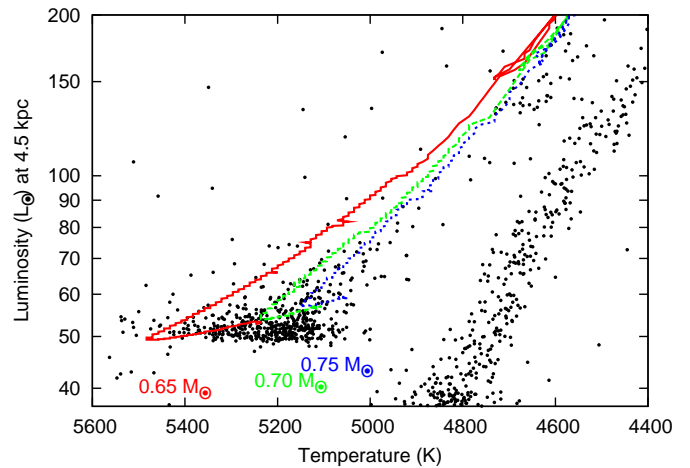}
\caption{As Fig.\ \ref{IsoDartmouthHB}, but with Pisa HB models at (left-to-right) 0.65, 0.70 and 0.75 M$_\odot$.}
\label{IsoPisaHB}
\end{figure}

\begin{figure}
\includegraphics[width=0.35\textwidth,angle=-90]{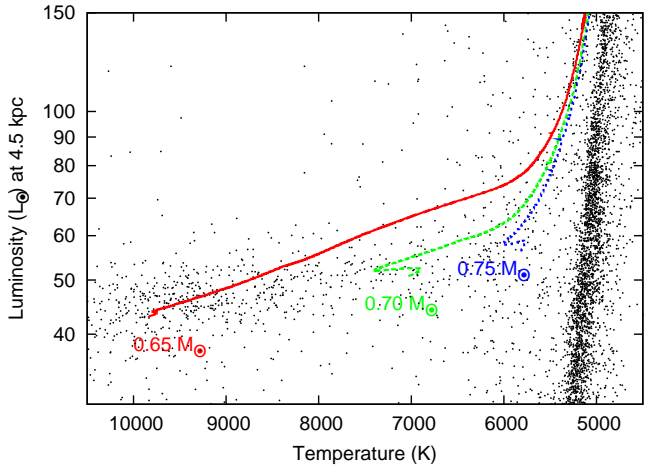}
\caption{As Fig.\ \ref{IsoPisaHB}, for the cluster $\omega$ Cen.}
\label{IsoPisaOmega}
\end{figure}

The amount of mass lost on the RGB can be determined by measuring the difference between the RGB-tip stars' initial masses and the mass of stars on the HB. The age of the stars means we can neglect mass loss during the comparatively short ($\sim$1.8 Myr; \citealt{SACWV10}) evolutionary period between the RGB-tip and zero-age HB. Initial masses for isochrones at 12 $\pm$ 1 Gyr are:
\begin{list}{\labelitemi}{\leftmargin=1em \itemsep=0pt}
\item $M_{\rm init}$ = 0.893 $^{+0.023}_{-0.021}$ M$_\odot$ for the Padova isochrones;
\item $M_{\rm init}$ = 0.874 $^{+0.022}_{-0.019}$ M$_\odot$ for the Dartmouth isochrones at [$\alpha$/Fe] = 0;
\item $M_{\rm init}$ = 0.889 $^{+0.025}_{-0.022}$ M$_\odot$ for the Dartmouth isochrones at [$\alpha$/Fe] = +0.2;
\item $M_{\rm init}$ = 0.916 M$_\odot$ for the Pisa evolutionary models (from \citealt{GCB+10}).
\end{list}

HB star masses can be estimated from synthetic HB models. The temperature spread of the HB is greater than that of the RGB at the same luminosity (the standard deviations from the mean temperature are $\approx 2.2\%$ \emph{vs.}\ $\approx 1.6\%$, respectively). This implies a moderate variation of stellar parameters among the HB stars. Notably, the HB lies at roughly constant luminosity, though luminosity may rise very slightly towards higher temperatures.

Figures \ref{IsoDartmouthHB} \& \ref{IsoPisaHB} show Dartmouth and Pisa synthetic HB models (\citealt{DCJ+08}\footnote{http://stellar.dartmouth.edu/~models/}; \citealt{CDIM+03}\footnote{http://astro.df.unipi.it/SAA/PEL/Z0.html}) for different stellar masses. The Dartmouth models are given at [$\alpha$/Fe] = 0 and +0.2; the Pisa models at $Y = 0.24$. Two things are immediately obvious from these figures. Firstly, that the temperature spread of the HB is not due to different stellar masses: the near-constant-luminosity morphology is more consistent with a spread in $\alpha$-element enhancement. Secondly, the HB star masses predicted by the two synthetic models varies considerably. Including the 2.6\% systematic luminosity uncertainty and 4.5\% distance uncertainty, we estimate from Figures \ref{IsoDartmouthHB} \& \ref{IsoPisaHB} that the mass of the HB stars is 0.90 $\pm$ 0.05 and 0.92 $\pm$ 0.05 M$_\odot$ for the Dartmouth models, assuming [$\alpha$/Fe] $\approx$ 0 and +0.2, respectively, and 0.68 $\pm$ 0.03 M$_\odot$ using the Pisa HB tracks.

This shows that the two HB models are clearly mutually incompatible, and that significant work in this area must be done before an agreement can be reached. The Dartmouth model suggests that RGB mass loss must be remarkably low, at $\lesssim 0.06$ M$_\odot$ (at the +1-$\sigma$ level, i.e.\ for 84\% of the stars). The Pisa models, however, predict RGB mass loss to be 0.24 $\pm$ 0.04 M$_\odot$ at the $\pm$1-$\sigma$ level (in agreement with the 0.242 M$_\odot$ found by \citet{GCB+10}). 

The small range of modelled HB-star masses provided by each set of models also suggests that mass loss \emph{on} the HB is low ($\lesssim 0.1$ M$_\odot$) in 47 Tuc, which would indicate that chromospherically-driven winds are not so important in this situation as implied by \citet{DSS09}: more consistency between the HB models would be necessary to substantiate this statement.

The integrated mass loss and median mass predicted by the Pisa models ($\approx$0.68 M$_\odot$) is in good agreement with that of the stars in $\omega$ Cen (Figure \ref{IsoPisaOmega}; data from Paper I). The majority of $\omega$ Cen's stars are 8--9 times more metal-poor than those of 47 Tuc, but crucially are of very similar age (hence similar initial masses). Due to the lack of consensus between the models, we use the Pisa models as a rough upper limit to the integrated RGB mass loss occurring in 47 Tuc.

It should also be possible to determine the mass of HB stars directly, by taking our photometrically-derived luminosities and temperatures, and comparing them to literature spectroscopically-derived surface gravities ($g$), via:
\begin{equation}
	R^2 = \frac{L}{4 \pi \sigma T^4} = \frac{G M}{g} ,
\label{MEqn}
\end{equation}
where the symbols take on their usual meanings. Due to the systematic uncertainties (in particular in measuring surface gravity), this can only be done with confidence on a set of stars reduced using identical processes, and preferably with similar temperature and luminosities. Unfortunately, there exist no spectroscopically-derived measurements of surface gravity measurements for a selection of both RGB and AGB stars with the internal accuracy required to do this.

If one assumes an integrated RGB mass-loss rate of 0.22 M$_\odot$ per star, comparable with the 0.20--0.25 M$_\odot$ derived for $\omega$ Centauri in Paper I, an AGB star will typically have a mass of $\sim$0.67 M$_\odot$. This implies that only $\sim$0.15 M$_\odot$ of its envelope mass remains, to be subsequently lost towards (or at) the AGB tip. We stress, however, that systematic uncertainties in the HB models mean that this cannot be proven from our data. We return to this subject of integrated RGB mass loss in the discussion.


\section{Mass loss and dust production}
\label{SectMdot}

Mass loss during the TP-AGB can give rise to dust production in the stellar outflow. The chemistry of the dust is determined by the amount of dredge-up which occurs: this defines whether carbon or oxygen is more numerous in the stellar atmosphere. Carbon and oxygen combine in the atmosphere to form CO. The remaining carbon or oxygen then goes on to define whether the dust is carbon-rich (primarily in the form of amorphous carbon) or oxygen-rich (primarily in the form of silicates). While a few carbon stars are found in globular clusters (e.g.\ \citealt{vLvLS+07}), they tend to be rare: their low mass means third dredge-up is not usually sufficient to cause the C/O ratio to exceed unity, thus most stars produce silicate dust. Metallic iron dust can also form around oxygen-rich stars \citep{KdKW+02,VvdZH+09,MSZ+10}. It is not clear whether metallic iron dust forms in carbon-rich stars too, due to the difficulty of spectroscopically separating metallic iron from amorphous carbon dust.

The dust can then drive a dusty wind by absorbing momentum from stellar radiation incident on it. This radiation pressure forces the dust from the star. Dust grains are collisionally-coupled to some extent with the surrounding gas, meaning that the gas is driven away from the star as well. It may be that stellar pulsations provide a net outward velocity to the dust, either as an initial `kick' velocity, or by the dissipation of acoustic energy in the extended atmosphere (e.g.\ \citealt{Bowen88,Lewis89,vLCO+08}).

Dust will re-radiate absorbed optical stellar light in the IR. This has the apparent effect of cooling the star and giving it excess in the IR above our model spectrum. Dust emits a modified blackbody spectrum which exists in addition to the stellar photospheric output at those wavelengths. Amorphous carbon, graphite and iron dust have no IR spectral features: such dust will typically give rise to positive values for the colours [5.8]--[8], [4.5]--[5.8] and even [3.6]--[4.5], though the latter depends on the dust's temperature. Particularly warm dust may also produce ($K_{s}$--[3.6]) $>$ 0 in addition to the above colours. Silicate dust leads to broad emission at 9.5 and 18 $\mu$m, which is partly covered by \emph{Spitzer's} 8- and 24-$\mu$m filters, giving rise to an excess in these bands.

\subsection{The infrared excess stars}
\label{XSStarsSect}

\subsubsection{Determining infrared excess}
\label{IRXSSect}

\begin{figure*}
\includegraphics[width=0.7\textwidth,angle=-90]{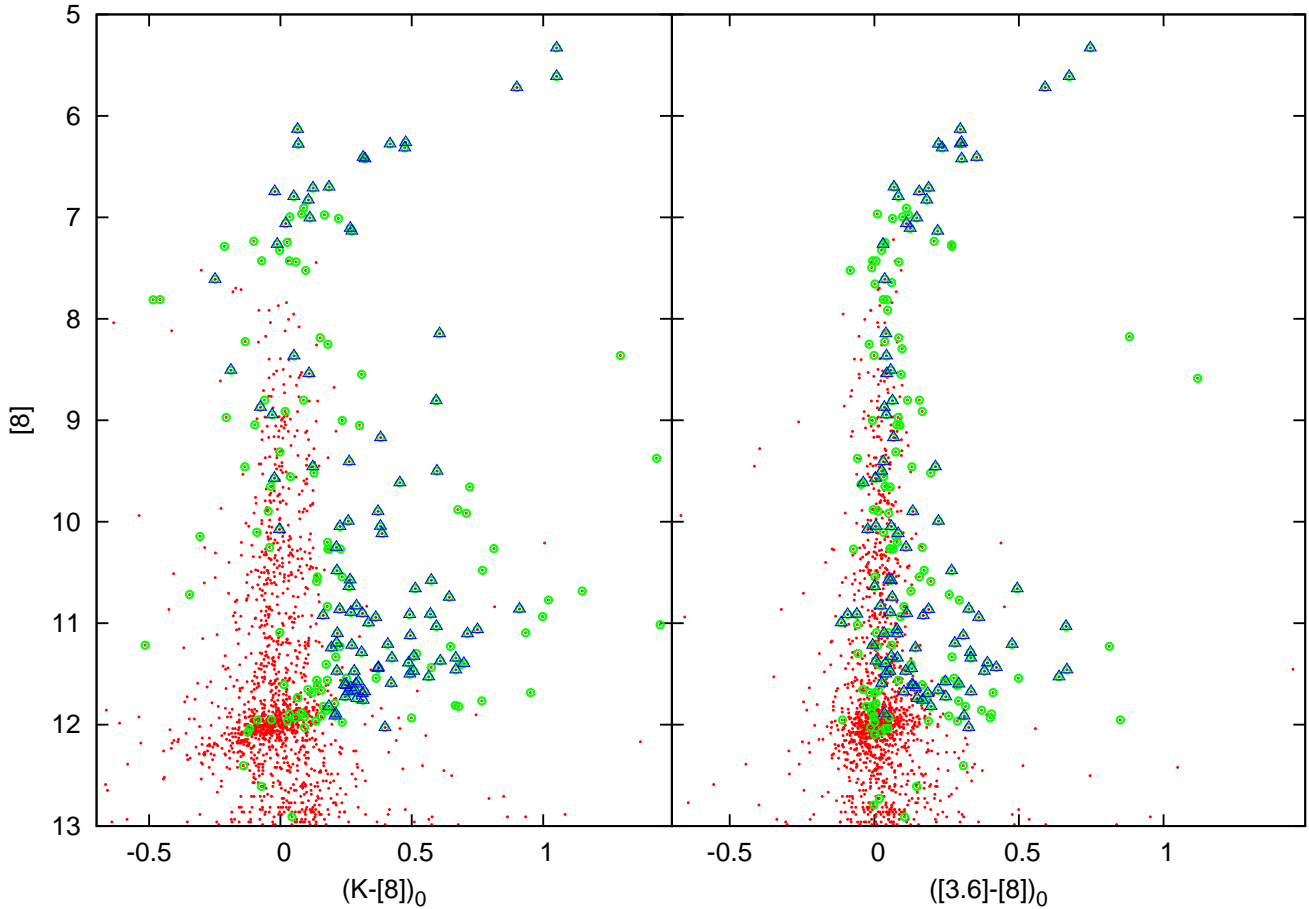}
\caption{Absolute $(K_{s}-[8])$ and $([3.6]-[8])$ colours for all stars in our sample (small, red dots). Circled dots (green) show our initial cut for dusty candidates. Triangles (blue) show our final cut for stars which we scrutinise individually, once bad data have been cleaned.}
\label{K8Fig}
\end{figure*}

Stars will exhibit IR excess for a variety of reasons not attributable to circumstellar dust. These include:
\begin{list}{\labelitemi}{\leftmargin=1em \itemsep=0pt}
\item Increased photometric and astrometric uncertainty due to problems with source separation in the dense cluster core;
\item Photometric errors due to `ghost' stars and linear artifacts near bright stars (caused by the bandwidth effect and banding, respectively\footnote{Details can be found in the IRAC handbook: http://ssc.spitzer.caltech.edu/irac/iracinstrumenthandbook/});
\item Artificial brightening due to blending with unresolved objects in mid-IR observations, which are at lower resolution than their near-IR/optical counterpart observations; and
\item Artificial brightening due to blending with other IR-bright objects, such as background galaxies or dusty SMC stars.
\end{list}
Of these, the first two will increase roughly as the square of the stellar density, while the latter two will increase proportionally with the stellar density.

To ensure we only select those stars that harbour circumstellar dust, we wish to perform a visual inspection of the photometry and imagery of each star. Such an inspection is performed to mitigate against data artifacts that may otherwise go undetected. It also allows us to sensibly include photometry from other literature sources (e.g.\ the \emph{AKARI} IRC point source catalogue; \citealt{KAC+10}) which improve coverage of dust features in the SED, but where coarser resolution and/or lower signal-to-noise mean flux measurements need to be interpreted with caution on an individual basis.

It would take prohibitively long to do this for our original list of 46\,398 stars. We reduce our source list to a manageable number by selecting only stars with $L > 31.6$ L$_\odot$ (i.e.\ the 3299 stars in or above the red clump) which have \emph{Spitzer} photometry (1993 of those stars) and meet one or more of the following criteria:
\begin{list}{\labelitemi}{\leftmargin=1em \itemsep=0pt}
\item is an evolved star candidate (\S\ref{HRSect}; 58 stars);
\item has at least two of the ($K_{s}$--[3.6]), ([3.6]--[5.8]), ([4.5]--[8]) or ([8]--[24]) colours which are more than 0.1 mag greater than that predicted by the model (74 stars);
\item has ($K_{s}$-[8]) or ([3.6]--[8]) at least 0.1 mag greater than that predicted by the model, if more-luminous than 60 L$_\odot$, or 0.2 mag greater otherwise (129 and 80 stars for each colour, respectively);
\item has ($K_{s}$-[24]) or ([3.6]--[24]) excess under the above criteria (36 and 42 stars for each colour, respectively);
\item has photometry from \emph{AKARI} (35 stars);
\item is a known variable (43 stars).
\end{list}
When combined, this list gives 258 unique objects. One may wonder why we use colours relative to our model in this case, rather than the absolute colours. In cooler stars, CO bands at 2.3 and 4.5 $\mu$m cause substantial deficits in the $K_{s}$ and 4.5-$\mu$m bands. In cool stars, around 3500 K, we model the colours of a naked star to be ($K_{s}$--[3.6]) = 0.17, ([3.6]--[5.8]) = 0.05, ([4.5]--[8]) = 0.3, ($K_{s}$-[8]) = 0.33 and ([3.6]--[8]) = 0.16 mag. These colours decrease in warmer stars. By taking into account the expected colours and their variation with stellar parameters, we can probe smaller excesses than possible with conventional colour--magnitude diagrams. The stars we investigate are shown in such a colour--magnitude diagram in Figure \ref{K8Fig}. 

A large number of stars were flagged where the photometry agreed between the 3.6- and 5.8-$\mu$m bands and between the 4.5- and 8-$\mu$m bands, but were discrepant between these two pairs. This is likely due to the simultaneity of the 3.6- and 5.8-$\mu$m, and 4.5- and 8-$\mu$m observations. It could either be a direct consequence (intrinsic variability) or indirect consequence (such as differing qualities of photometry due to different coverage in the original mosaics). For this reason, we have ignored the ([3.6]--[4.5]), ([4.5]--[5.8]) and ([5.8]--[8]) colours in the above criteria.

\begin{center}
\begin{table}
\caption{Sources showing IR excess in both \protect\cite{ORFF+07} and this work.}
\label{OrigliaTable}
\begin{tabular}{c@{\ \ \ }c@{\ \ }c@{\ \ }c@{\ \ }l@{}}
    \hline \hline
ID$^1$& Offset$^2$		& Temp.	& Lum. & Notes$^3$\\
   	& ($^{\prime\prime}$)	& (K)		& L$_\odot$  & \\
    \hline
1	& 0.21	& 3623	& 4824 & V1\\
8	& 0.33	& 3578	& 3583 & V8\\
4	& 0.20	& 3521	& 2603 & V4\\
3	& 0.74	& 3500	& 2541 & LW13\\
5	& 0.59	& 3543	& 2325 & LW10\\
2	& 0.26	& 3575	& 2301 & V21\\
7	& 0.23	& 3374	& 2204 & LW9\\
6	& 0.22	& 3510	& 2140 & V27\\
9	& 0.25	& 3526	& 2096 & A19\\
12	& 0.17	& 3713	& 2079 & LW12\\
13	& 0.30	& 3591	& 1941 & FBV45\\
10	& 0.16	& 3644	& 1822 & LW18\\
18	& 0.17	& 3802	& 1773 & LW15\\
26	& 0.30	& 3816	& 1640 & MVx03\\
19	& 0.19	& 3738	& 1638 & LW19\\
21	& 0.21	& 3687	& 1635 & LW1\\
14	& 0.20	& 3602	& 1575 & V20\\
34	& 0.16	& 3935	& 1443 & LW6\\
44	& 0.08	& 3763	& 1374 & V6\\
39	& 0.63	& 3915	& 1293 & LW8\\
30	& 0.14	& 3782	& 1274 & LW7\\
36	& 0.12	& 3976	& 1142 & 00$^h$24$^m$06.3$^{\prime\prime}$ --72$^\circ$04$^\prime$45$^{\prime\prime}$\\
33	& 0.15	& 3875	& 1114 & 3$^{\prime\prime}$ W of LW16\\
52	& 0.13	& 4096	& 940  & 00$^h$24$^m$01.7$^{\prime\prime}$ --72$^\circ$04$^\prime$56$^{\prime\prime}$\\
67	& 0.31	& 4124	& 699  & 00$^h$24$^m$05.1$^{\prime\prime}$ --72$^\circ$04$^\prime$54$^{\prime\prime}$\\
86	& 0.38	& 4138	& 579  & 00$^h$24$^m$07.7$^{\prime\prime}$ --72$^\circ$05$^\prime$20$^{\prime\prime}$\\
81	& 0.52	& 4067	& 561  & 00$^h$24$^m$08.4$^{\prime\prime}$ --72$^\circ$04$^\prime$14$^{\prime\prime}$\\
99	& 0.52	& 4105	& 477  & 00$^h$23$^m$56.2$^{\prime\prime}$ --72$^\circ$04$^\prime$47$^{\prime\prime}$\\
171	& 0.22	& 4203	& 441  & 00$^h$24$^m$28.9$^{\prime\prime}$ --72$^\circ$04$^\prime$43$^{\prime\prime}$\\
138	& 0.03	& 4186	& 334  & R13\\
148	& 1.11	& 4079	& 334  & FBV42\\
181	& 0.29	& 4487	& 326  & 00$^h$24$^m$10.7$^{\prime\prime}$ --72$^\circ$04$^\prime$41$^{\prime\prime}$\\
222	& 0.11	& 4591	& 306  & FBV35\\
189	& 0.46	& 4447	& 282  & R7\\
195	& 0.18	& 4389	& 264  & FBV34\\
106	& 0.94	& 2730	& 237  & 00$^h$24$^m$15.4$^{\prime\prime}$ --72$^\circ$05$^\prime$06$^{\prime\prime}$\\
235	& 0.54	& 4275	& 237  & 00$^h$24$^m$06.9$^{\prime\prime}$ --72$^\circ$05$^\prime$21$^{\prime\prime}$\\

373 	& 0.46 	& 4723 	& 215  & 00$^h$24$^m$10.5$^{\prime\prime}$ --72$^\circ$07$^\prime$16$^{\prime\prime}$\\
224 	& 0.24 	& 4491 	& 187  & 00$^h$23$^m$53.1$^{\prime\prime}$ --72$^\circ$05$^\prime$10$^{\prime\prime}$\\
309 	& 0.50 	& 4550 	& 178  & 00$^h$24$^m$00.6$^{\prime\prime}$ --72$^\circ$04$^\prime$47$^{\prime\prime}$\\
356 	& 0.25 	& 4946 	& 177  & 00$^h$24$^m$07.1$^{\prime\prime}$ --72$^\circ$05$^\prime$01$^{\prime\prime}$\\
479 	& 1.62 	& 4201 	& 170  & 00$^h$24$^m$12.4$^{\prime\prime}$ --72$^\circ$06$^\prime$07$^{\prime\prime}$\\
296 	& 0.44 	& 4457 	& 161  & 00$^h$24$^m$03.3$^{\prime\prime}$ --72$^\circ$04$^\prime$31$^{\prime\prime}$\\
345 	& 0.10	& 4632 	& 152  & 00$^h$23$^m$55.9$^{\prime\prime}$ --72$^\circ$04$^\prime$54$^{\prime\prime}$\\
428 	& 0.19 	& 4618 	& 112  & 00$^h$24$^m$12.6$^{\prime\prime}$ --72$^\circ$04$^\prime$24$^{\prime\prime}$\\
    \hline
\multicolumn{5}{p{0.47\textwidth}}{$^1$ID number from \protect\citet{ORFF+07}. $^2$Offset in arcseconds between \citet{ORFF+07} and this work, limited to $<$2$^{\prime\prime}$. $^3$ Variable (V, LW, A) names from \citet{Clement97}; and \citet{LW05}. Other names from: Lee --- \citet{Lee77}; FBV --- \citet{FBV+02}; MV --- \citet{MvL07}; R --- \citet{FT60}.}\\
    \hline
\end{tabular}
\end{table}
\end{center}

\begin{figure}
\centerline{\includegraphics[width=0.35\textwidth,angle=-90]{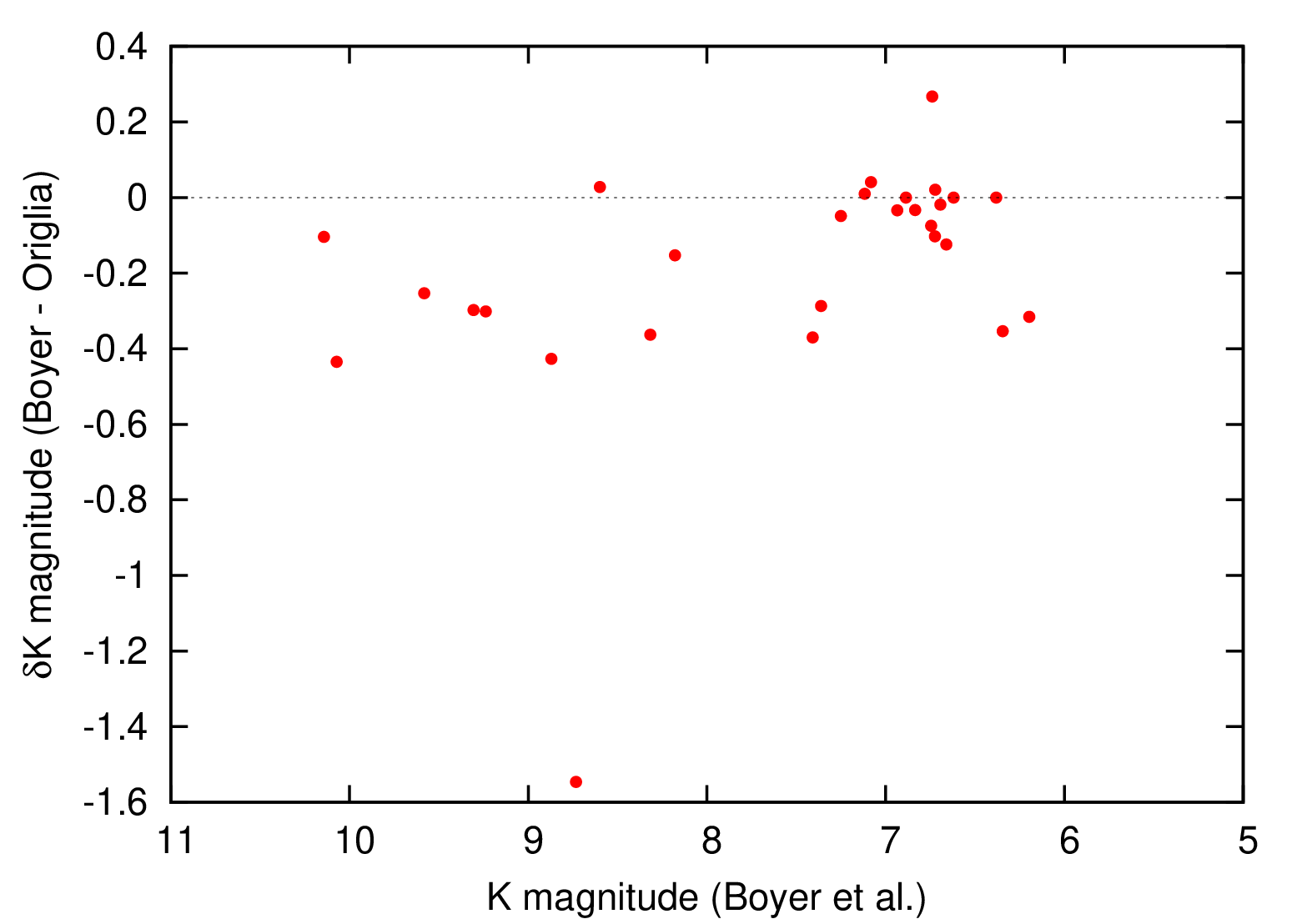}}
\centerline{\includegraphics[width=0.35\textwidth,angle=-90]{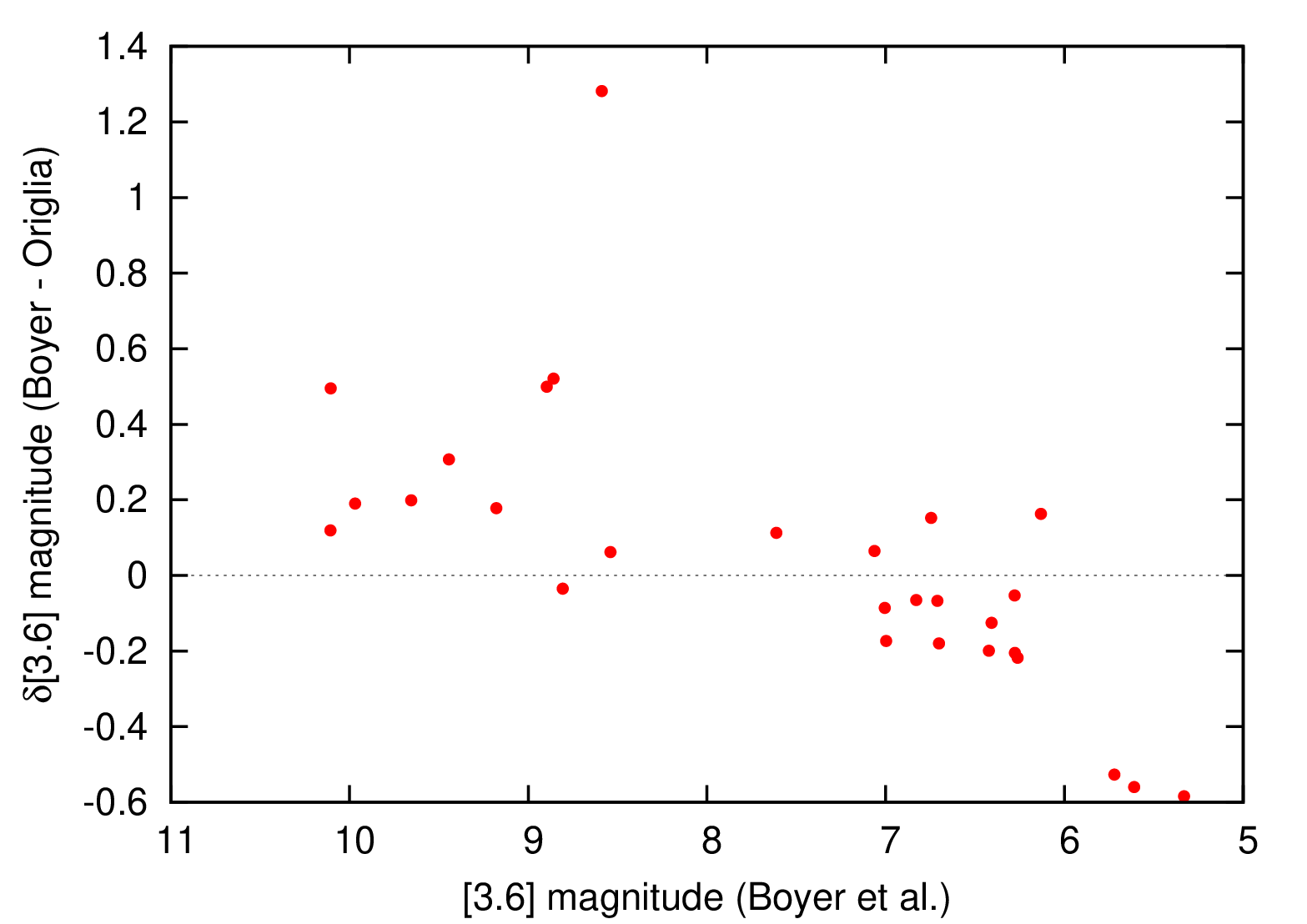}}
\centerline{\includegraphics[width=0.35\textwidth,angle=-90]{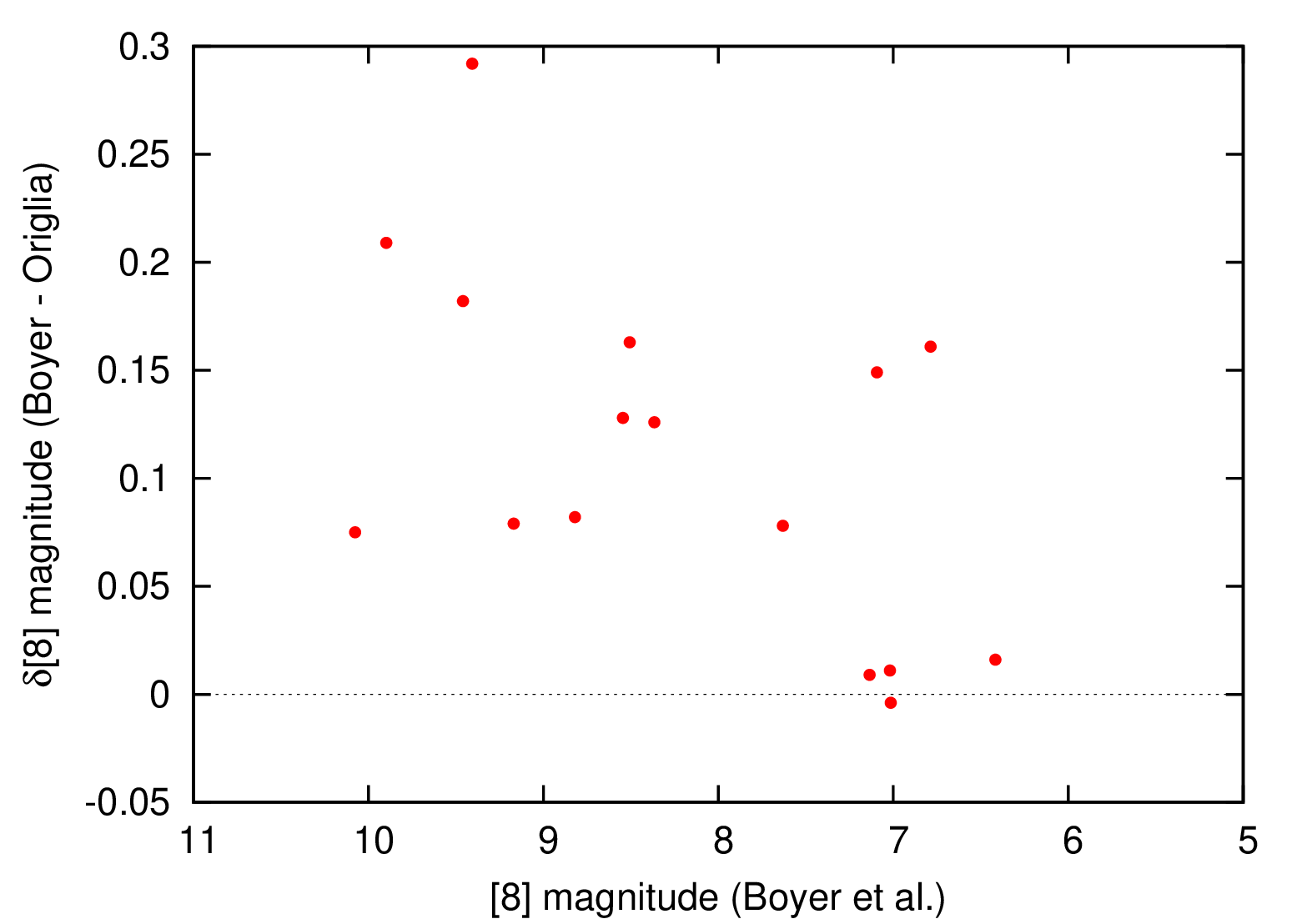}}
\caption{Differences in photometric measurements presented in \protect\citet{ORFF+07} and \protect\citet{BvLM+10}, showing only those stars suggested to harbour circumstellar dust by Origlia et al.}
\label{OrigCompFig}
\end{figure}

At this point, we have a list of potentially dusty objects, using criteria which are qualitatively similar to those applied in \citet{ORFF+07}. Astrometry and photometry of the sources in that work have been made available to us (L.\ Origlia, private communication). Of their 93 sources, only 45 are found to have counterparts with IR excess in our own candidate list (Table \ref{OrigliaTable}). This immediately suggests that the primary difference between the two works lies in the original photometric reduction, and that the real photometric errors are considerably larger than those used for the 3$\sigma$ cutoff in \citet{ORFF+07}. Of these 45 matched sources, 23 are over 1000 L$_\odot$, while only 22 are below 1000 L$_\odot$. We explicitly do not claim one reduction to be more accurate than the other: while several stars known to harbour dust do not appear in Origlia et al.'s list (e.g.\ V13 and V18; \citealt{vLMO+06,LPH+06}), they lie outside their survey region.

One comparison we can make is of the magnitudes of Origlia et al.'s dusty stars used in their work, and those used in this work (Figure \ref{OrigCompFig}). It is notable that (particularly for the fainter stars) the $K_{s}$-band magnitudes are systematically brighter in this work, especially where lower-resolution 2MASS data are available, while the [3.6] and [8] magnitudes are systematically fainter. This is further evidence that the method of photometric reduction and differences in resolution are the major differences between \citet{ORFF+07} and \citet{BvLM+10}.

\subsubsection{Sample cleaning}
\label{CleaningSect}

\begin{figure}
\centerline{\includegraphics[height=0.50\textwidth,angle=-90]{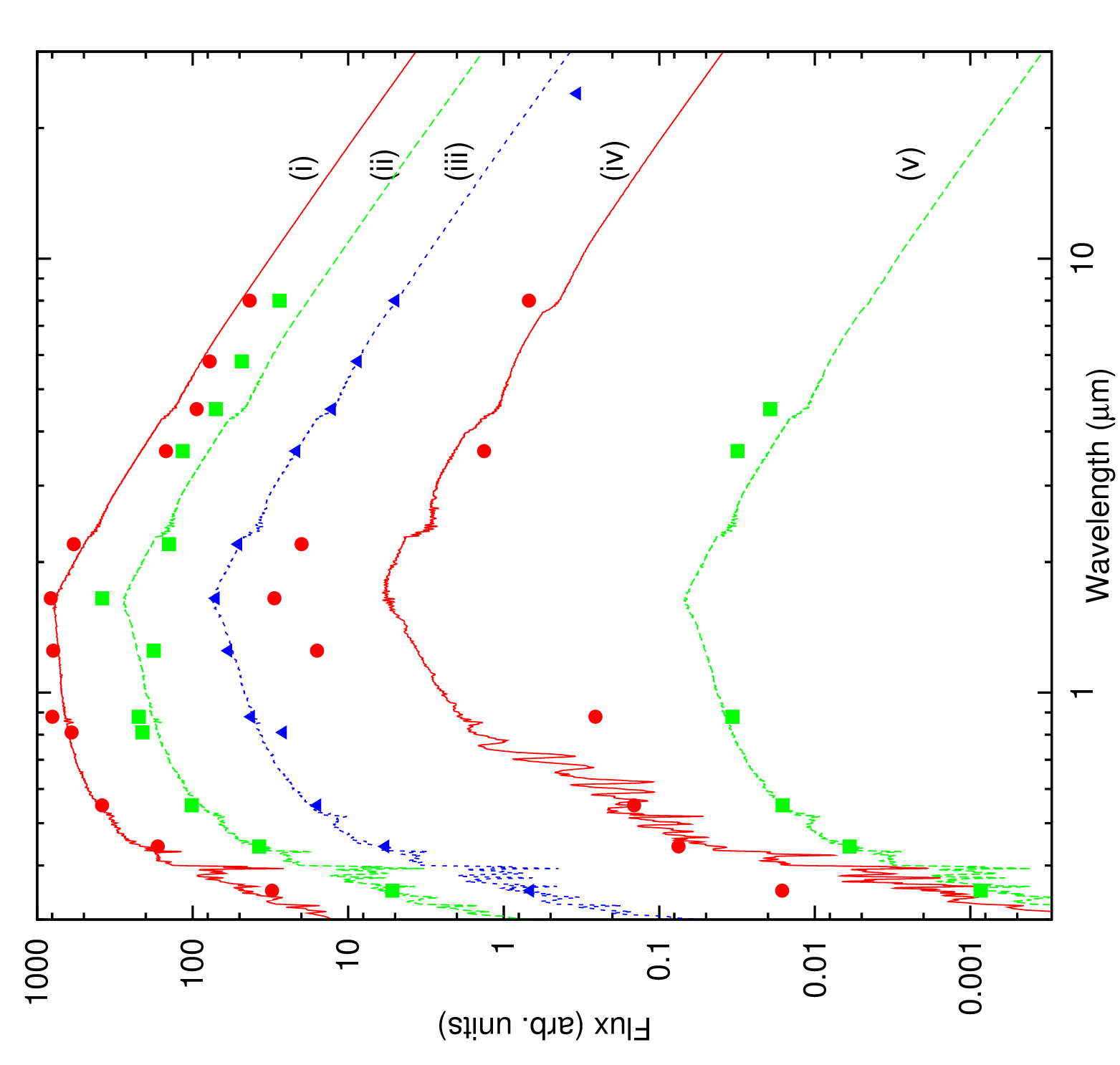}}
\caption{Examples of stars rejected as having poor photometry. See \S\protect\ref{CleaningSect} for details.}
\label{RejectsFig}
\end{figure}

Most of the 258 objects that match our criteria above were flagged as a result of incorrectly-matched photometry, either between surveys in our cross-matching routines, or within the individual surveys themselves. The Gunn $i$-band data from the MCPS and the $JHK_{s}$ data from 2MASS are particularly prone to these errors. We expect dusty stars to have a mid-IR excess, and that the ratio of observed/modelled flux should not appreciably decrease toward longer wavelengths. To remove stars which have been erroneously selected due to bad photometric data or matching, we have removed stars selected with the following cuts:
\begin{list}{\labelitemi}{\leftmargin=1em \itemsep=0pt}
\item To remove objects with poor $JHK_{s}$-\emph{Spitzer} matching, we remove objects which show $K_{s}$-band excesses (compared to our model) more than 0.1 mag greater than the associated 3.6-$\mu$m excess. An example is shown in Figure \ref{RejectsFig}, (i). This reduces our source list to 221 objects.
\item To remove objects with poor $JHK_{s}$ matching in general, we remove objects which have observed/model flux ratios at $J$, $H$ and $K_{s}$-band of $<$0.8 (i.e.\ a strong near-IR flux deficit; Figure \ref{RejectsFig}, (ii)). This leaves 188 objects.
\item To remove objects with bad photometry in \emph{Spitzer} data, while retaining stars with only 8- or 24-$\mu$m excess, we remove objects which show less than 10\% excess above the model at the longest recorded wavelength: 12 sources are discounted because they show no excess at 24 $\mu$m and 37 because they show no excess at 8 $\mu$m (when 24 and 8 $\mu$m are the longest wavelength with data, respectively). These stars mostly have discordant IRAC photometry, but include V16, LW1, LW2 and LW21, which were selected due to their variability, but show no apparent mid-IR excess (Figure \ref{RejectsFig}, (iii)). This leaves 128 objects.
\item To remove objects with bad photometric data at $K_{s}$-band, we remove objects which show an observed/model flux ratio at $K_{s}$-band which is $>$0.1 mag less than that at both $J$-band and $H$-band (Figure \ref{RejectsFig}, (iv); leaving 127 objects).
\item To remove objects with mid-IR excess caused by missing near-IR data (leading to unconstrained fits), we remove objects with no $K_{s}$-band data unless ([3.6]--[8]) $> 0.1$ mag (Figure \ref{RejectsFig}, (v); leaving 120 objects).
\end{list}

This reduction removes most of the faint giants from the sample, but also a few bright giants. These include V19, LW1, and the blended sources LW7 and LW8. Comparing our reduced sample to \citet{ORFF+07} shows that, out of these 120 sources, 28 match theirs. Of these 28, 18 are above 1000 L$_\odot$ and ten are below 1000 L$_\odot$. The ten objects are listed in \citet{ORFF+07} as IDs 67, 81, 86, 148 (FBV42), 181, 189 (R7), 222 (FBV35), 296, 309 and 373. We note that the study by \citet{FBV+02}, whose authors lend their names to the FBV stars, shows no notably-different polarisation (which would be suggestive of additional, circumstellar, line-of-sight dust) in either FBV35 or FBV42, in comparison to the brighter, variable stars.

As both studies find a significant fraction of potentially dust-harbouring stars in the core, we may expect some overlap between the two samples, even if they are randomly distributed. The probability ($P$) that two samples of stars ($m$ and $n$) are randomly drawn from a total of $N$ stars when $i$ stars are found in both samples can be found (in the limit of identical object detection efficiency and independent datasets) by:
\begin{equation}
P = 1 - \sum_{k=0}^{i}
	\left(\frac{\left(\frac{m!}{k!(m-k)!}\right)
			\left(\frac{(N-m)!}{(n-k)!\ ([N-n]-[m-k])!}\right)}
	{\left(\frac{N!}{n!(N-n)!}\right)}\right) .
\label{HypergeoEqn}
\end{equation}

For the entire sample of stars, $N$ = 3399, $n$ = 103 (not 120, as 17 stars are identified outside the region covered by Origlia et al.), $m$ = 93 and $k$ = 28, giving an answer of $P \ll 10^{-10}$: i.e.\ zero chance of the overlap being due to chance. Concentrating on the cluster's inner 2$^\prime$, however, we find $P = 0.01$\% for the bright giants\footnote{$N$ = 44, $n$ = 21, $m$ = 22, $k$ = 16} ($[8]<8$, $L \gtrsim 1000$ L$_\odot$) and $P = 8.5$\% for the faint giants\footnote{$N$ = 254, $n$ = 29, $m$ = 65, $k$ = 10} ($8<[8]<10.874$, $35 \lesssim L \lesssim 1000$ L$_\odot$). Considering only the inner 1$^\prime$, $P = 0.15$\% for the bright giants\footnote{$N$ = 24, $n$ = 15, $m$ = 10, $k$ = 9} and $P = 22.2$\% for the faint giants\footnote{$N$ = 131, $n$ = 17, $m$ = 47, $k$ = 7}.

We can therefore state that the detection of infrared excess around the bright giants has a negligibly-low probability of being due to random selection (at a level of around 3--4$\sigma$). One may also conclude that the detection of infrared excess around the faint giants has a non-negligible probability of being taken from a random selection (at 22.2\%, or 1--2$\sigma$). However, we remind the reader that the two samples are taken from the same set of \emph{Spitzer} data and thus not independent: they will therefore have identical noise and image artifacts to contend with. The real probability that the two samples of stars trace objects showing a physical phenomenon is (unquantifiably) lower than the 77.8\% that the above probabilities predict. We are therefore not confident that the overlap of 28 stars between our two samples is significant, nor that our two studies are actually tracing a particular sub-population of stars.

\subsubsection{Image artifacts}

\begin{figure}
\includegraphics[width=0.47\textwidth]{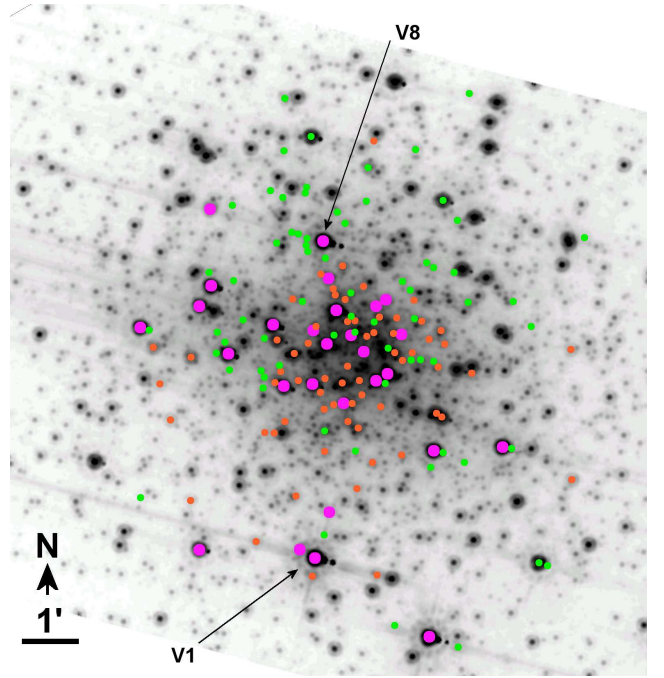}
\caption{\protect\emph{Spitzer} 8-$\mu$m imaging of the core of 47 Tuc from the Rood observations. Orange dots mark the stars found to have IR excess by \protect\citet{ORFF+07}; green circles mark our own dusty candidates; large magenta dots denote objects found to have IR excess by both parties. Banding is present to the west-north-west of bright stars: the brightest, V1 and V8, are labelled.}
\label{BandingFig}
\end{figure}

As noted in \S\ref{IRXSSect}, the original \emph{Spitzer} data used both in \citet{ORFF+07} and this work contain a number of known artifacts: notably banding, the bandwidth effect, and blending. Figure \ref{BandingFig} shows the central region of the cluster, containing almost all the stars suggested by \citet{ORFF+07} to contain dust, and also the vast majority of our 120 dusty candidates. From this image, the lack of overlap between our two candidate sets is clear. We now investigate how each of the effects affect our data, but stress that we do not discount any individual targets in this section.

Banding is caused by optical scattering and charge transfer within the IRAC array, which causes horizontal and vertical spikes to become manifest around bright stars. In Figure \ref{BandingFig}, these trails can be seen extending in the ``10-o$^\prime$clock'' and ``4--o$^\prime$clock'' directions from the brightest stars, with less-prominent lines perpendicular to this axis.

Banding has clearly affected both selections of dusty candidates: several candidates from \citet{ORFF+07} lie on banding, caused by V1 (IDs 195, 222, 293 and 379) amongst others; several of our candidates lie on banding caused by V8 (Figure \ref{BandingFig}). Of the stars both studies find to have IR excess, ID-148 and ID-222 are obviously affected by banding, though several other stars to the west of the cluster may also be affected by this problem (IDs 6, 9, 10, 19, 171) as may stars near the center of the cluster (including IDs 3, 26, 36, 67 and 181). While banding affects stars of all magnitudes, the amplitude of the effect scales inversely with received stellar flux. Banding can therefore not reproduce the excess in the brighter giants, whose excess is also confirmed by either their ([3.6]--[8]) colour and/or literature photometry (see below, \S\ref{BrightXSSect}).

Ghost images, caused by the bandwidth effect, affect the brightest stars at 8 $\mu$m. While this does not appear to have affected Origlia et al.'s selection, several of our dusty candidates lie on these ghost images, which artificially boosts their 8-$\mu$m flux, mimicking excess.

Blending is obviously very significant in the crowded cluster core, affecting a set of stars which include the aforementioned IDs 3, 26, 36, 67 and 181. Every star is blended to some extent, but it is difficult to quantify the amount of blending present. \citet{ORF+10} present strong evidence from \emph{Hubble Space Telescope} $I$-band imagery that very close blends with stars of similar $I$-band magnitude is not an issue in the majority of cases. What was not addressed are blends with objects of large ($I$--[3.6]) or ($I$--[8]) colours, and objects blended with other objects at distances of more than 1$^{\prime\prime}$.

The majority of very red sources, showing large ($I$--[3.6]) and ($I$--[8]) colours are galaxies (Paper I). We do not, however, consider this a major source of blending in this case, as these should affect stars randomly, rather than being more prevalent in the inner regions, and are likely to only affect one or two sources in the cluster (cf.\ Paper I; \citealt{MMN+08}).

On the other hand, blending with objects outside the central PSF is still highly significant, as in our example in \S\ref{K8Sect}, 2MASS\,00234588--7204488. There are two main reasons for this. First of all, these images are limited by diffraction, rather than seeing, meaning that a complex PSF extends a long way from the central source. If the flux ratio between the two blended sources is high enough, a small error removing flux from the brighter object can leave significant flux in the Airy rings it creates. Some of this flux can be attributed to the fainter object, causing it to be artificially brightened, and leading to an obviously non-stellar SED. 

A fundamental difference exists in the reduction process between the Origlia et al.\ papers and ours. Our {\sc daomatch/daomaster} reduction fits and removes the PSF of the brightest sources, then subsequently removes fainter objects until we have removed all significant flux from the image. Origlia et al.'s {\sc romafot}-based reduction fits PSFs to the lower-resolution, longer-wavelength \emph{Spitzer} data. They then sum the flux from all sources within their $K$-band data which lie inside the Airy radius of each \emph{Spitzer} source (2.3$^{\prime\prime}$ at 8 $\mu$m) and use Kurucz model atmospheres to determine the expected flux at longer wavelengths: stars with flux in excess to these values are determined to have infrared excess. Each technique has its benefits. Our more-conventional technique is optimised where the $K$-band and \emph{Spitzer} data have similar PSFs (as is the case with the 2MASS data), while Origlia et al.'s more-innovative approach should theoretically work better for their higher-resolution $K$-band data.

Our concern is whether their method is sophisticated enough to take into account all forms of blending: by summing flux from sources with centroids within 2.3$^{\prime\prime}$, a blended source at 2.2$^{\prime\prime}$ from a star is treated the same as one lying much closer to the star, while a source at 2.4$^{\prime\prime}$ from a star is not considered to affect the star at all. In the former case, the blending star will not contribute as much to the PSF as is calculated, leading to an over-estimation of the expected 8-$\mu$m flux; in the latter case, the blending star will contribute some unaccounted flux towards the PSF and lead to an under-estimation of the expected 8-$\mu$m flux.

This is of particular concern when a source lies on the Airy ring of a bright star, such as Origlia's ID-373, which is affected by the nearby V1. One can calculate the contamination between two stars by convolving the IRAC 8-$\mu$m PSF of the measured star, with the PSF of the blending star which has been shifted appropriately in the RA--Dec plane and multiplied by the fractional error in its flux. One can consider a typical example of two stars, where a bright giant is 2.4$^{\prime\prime}$ from a star 33$\times$ (3.8 mag) fainter than it. The bright giant will have a typical error in flux of 3\%, which corresponds to a flux comparable to that received from the fainter star. If one fits a PSF to the fainter star, one will find it 0.12--0.14 mag brighter than it actually is (depending on the position angle of the blended source). It is worth remembering that this is caused by a 1$\sigma$ error: with $\sim$20 giants bright enough to cause this effect, one may expect errors up to roughly twice this value. When one takes into account blending by multiple brighter objects (as happens in the cluster core) and any noise inherent in the data, this effect can increase further.

The second problem is that source confusion can lead to astrometric errors. A second fundamental difference exists between our reduction processes: we first fit PSFs to objects detected in each band, then combine the data from different bands together, while Origlia et al.\ fit PSFs to identical co-ordinates across all bands simultaneously. Again, each technique has its benefits: Origlia et al.'s method mitigates against displacement of the PSF centroid due to blended sources, while our technique avoids problems with image distortion and alignment within surveys, and co-ordinate system alignment among surveys.

While the co-ordinates of most of the brightest stars agree to within $\sim$0.2$^{\prime\prime}$ with the positions provided by Livia Origlia (private communication), LW10 and V26 are misplaced by 0.59$^{\prime\prime}$ and 0.74$^{\prime\prime}$, respectively. Both stars are $\approx$2500 L$_\odot$, meaning the poor removal of the stars' PSFs caused by such an error may lead to significant errors in other stars' fluxes. Both studies identify several (different) objects around both of these stars as having IR excess, but it is not clear that this is actually the case.

Despite the differences between our reduction methods, one would expect that if fainter stars had infrared excess created by dust, we would mostly identify the same stars as having infrared excess.

\subsubsection{Is $(K_{s}-[8])$ better than $([3.6]-[8])$?}
\label{K8Sect}

\begin{figure*}
\centerline{\includegraphics[width=0.35\textwidth,angle=-90]{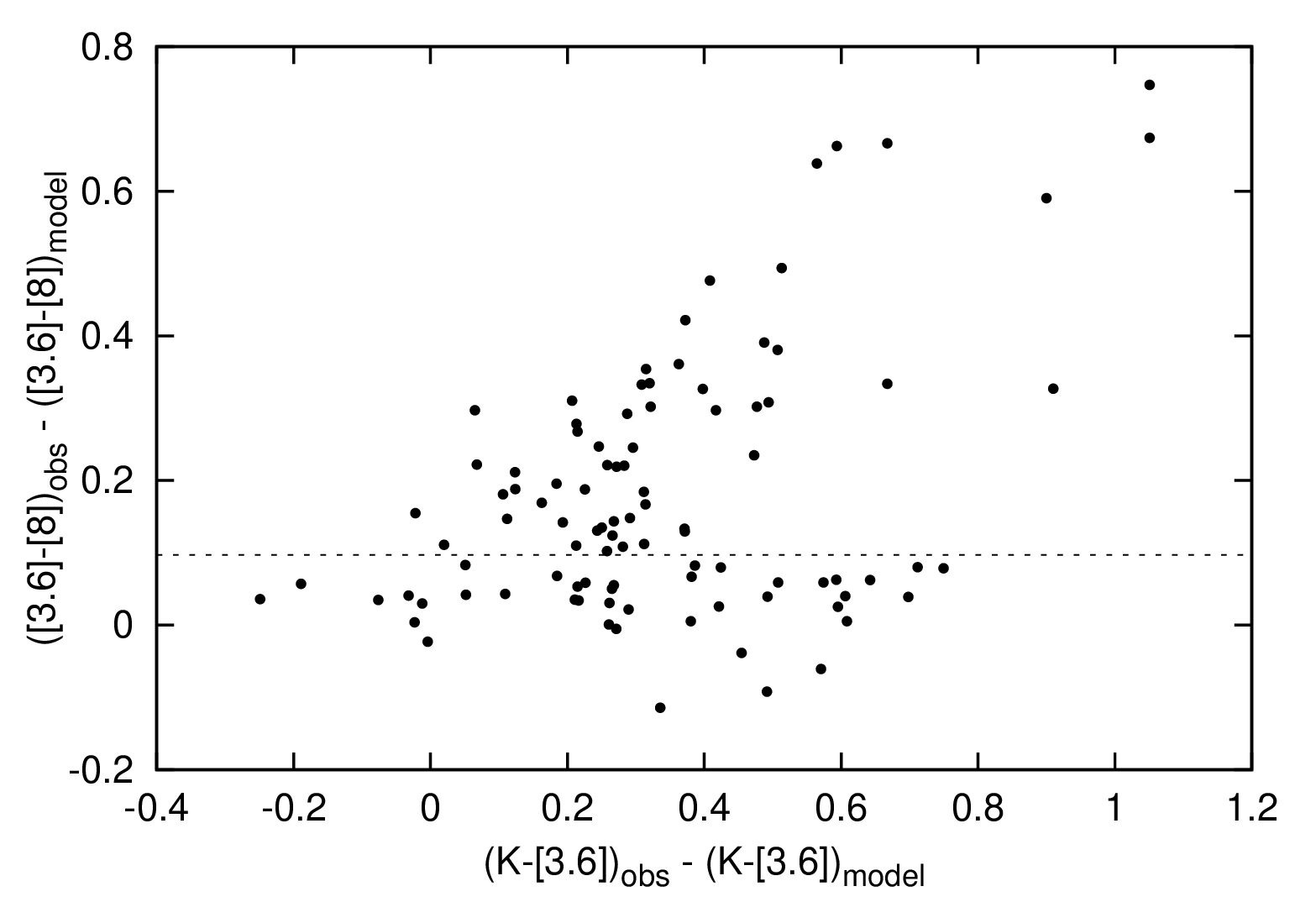}\includegraphics[width=0.35\textwidth,angle=-90]{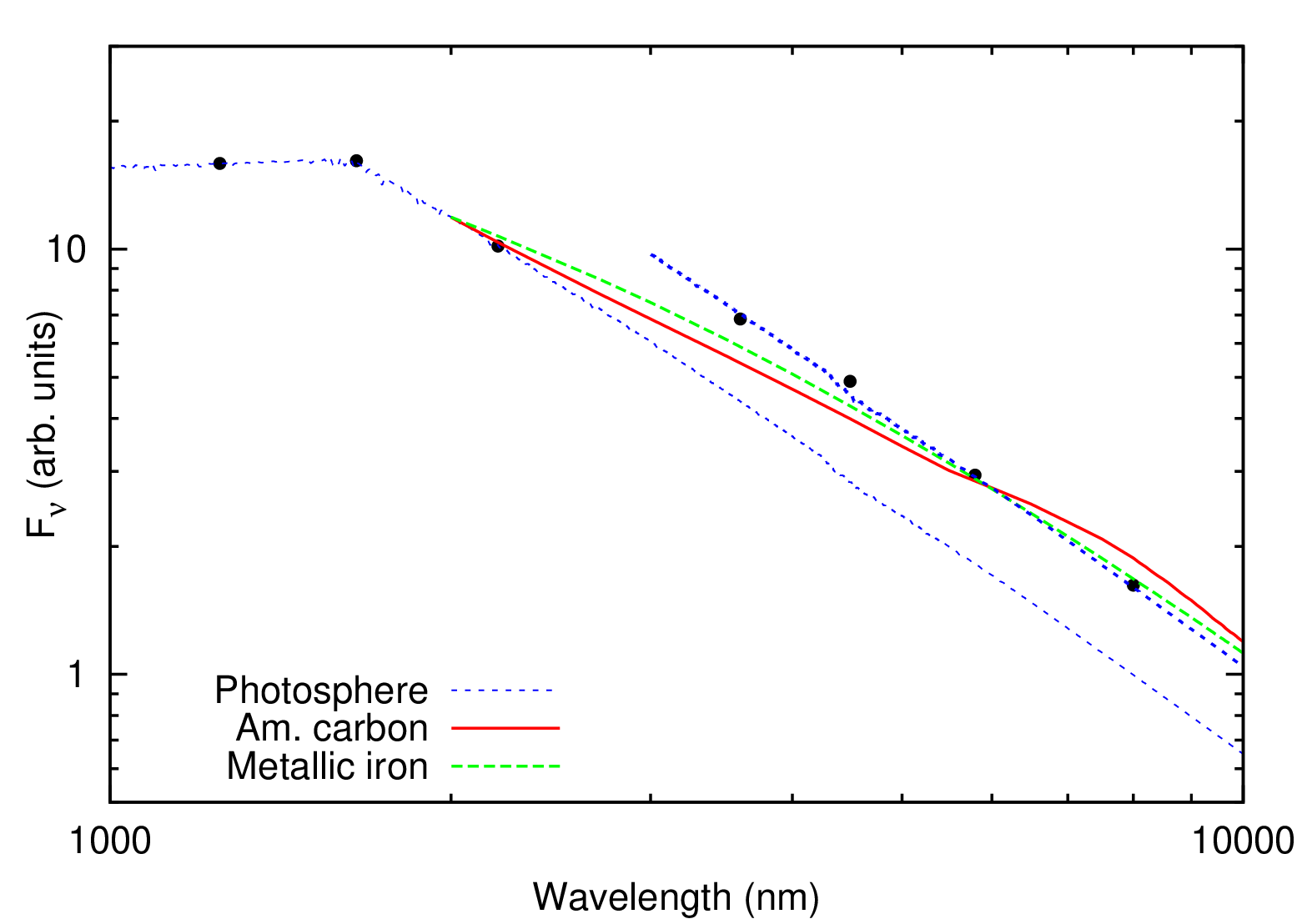}}
\centerline{\includegraphics[width=0.35\textwidth,angle=-90]{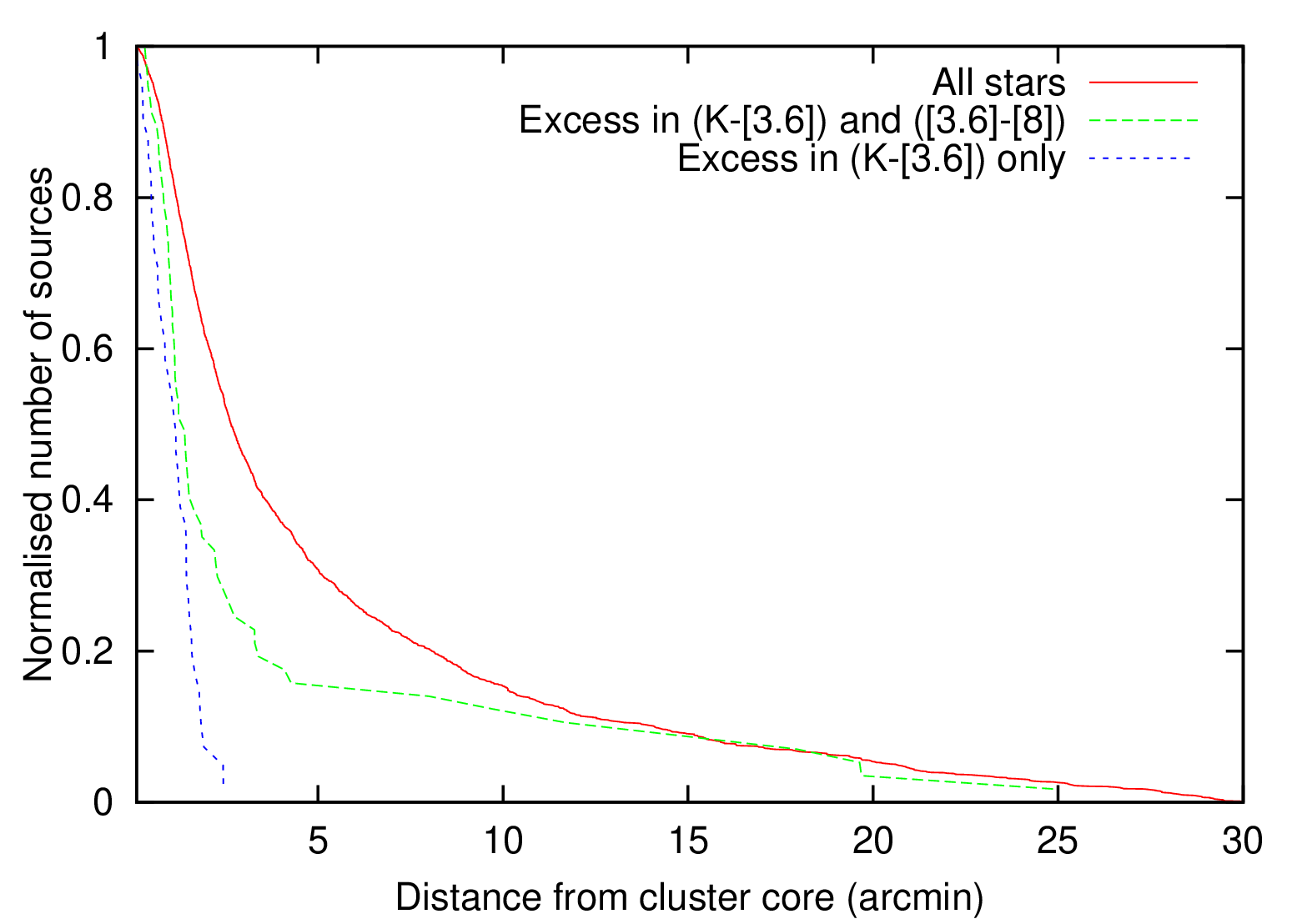}\includegraphics[width=0.35\textwidth,angle=-90]{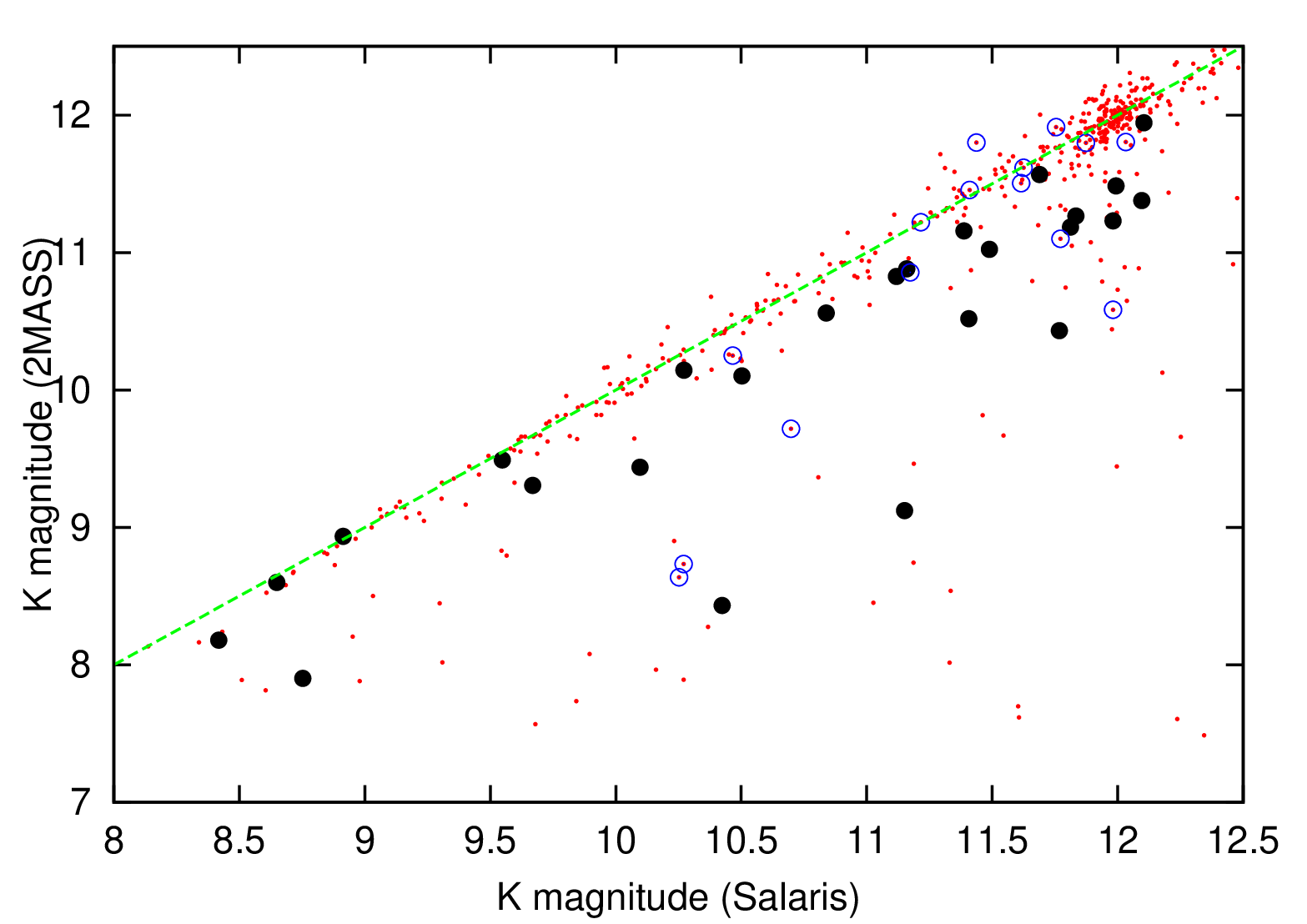}}
\caption{Separation of dusty stars from bad cross-matches and blends. Top-left panel: $(K_{s}-[3.6])$ and $([3.6]-[8])$ colours for all stars in our sample with respect to our model and our colour cut for separating the two populations. \\
Top-right panel: SED of source 5.9408750--72.0803889 (black points) which is a poorly-cross-matched source, showing: a model photosphere (short-dashed, blue lines) at the fitted luminosity and at 67\% brighter than the fitted luminosity, an amorphous carbon dust model at 1500 K (solid, red line), and a metallic iron dust model at 1500 K (long-dashed, green line). \\ 
Bottom-left panel: radial distribution of (solid, red line) all stars with $L > 31.6$ L$_\odot$ with \emph{Spitzer} photometry, (long-dashed, green line) stars with positive colours for both $(K_{s}-[8])$ and $([3.6]-[8])$, (short-dashed, blue line) stars with positive $(K_{s}-[8])$ colour but zero $([3.6]-[8])$ colour. \\ 
Bottom-right panel: $K_{s}$-band magnitudes of stars covered by both \protect\citet{SHO+07} and 2MASS. Dusty candidates below the line in the upper-left panel are shown as large, black dots, while those above the line are shown as large, blue circles; the line denotes parity between surveys. (Note: the brightest stars are not present in the Salaris data due to saturation problems.)}
\label{RealDustFig}
\end{figure*}

\begin{figure}
\centerline{\includegraphics[width=0.35\textwidth,angle=-90]{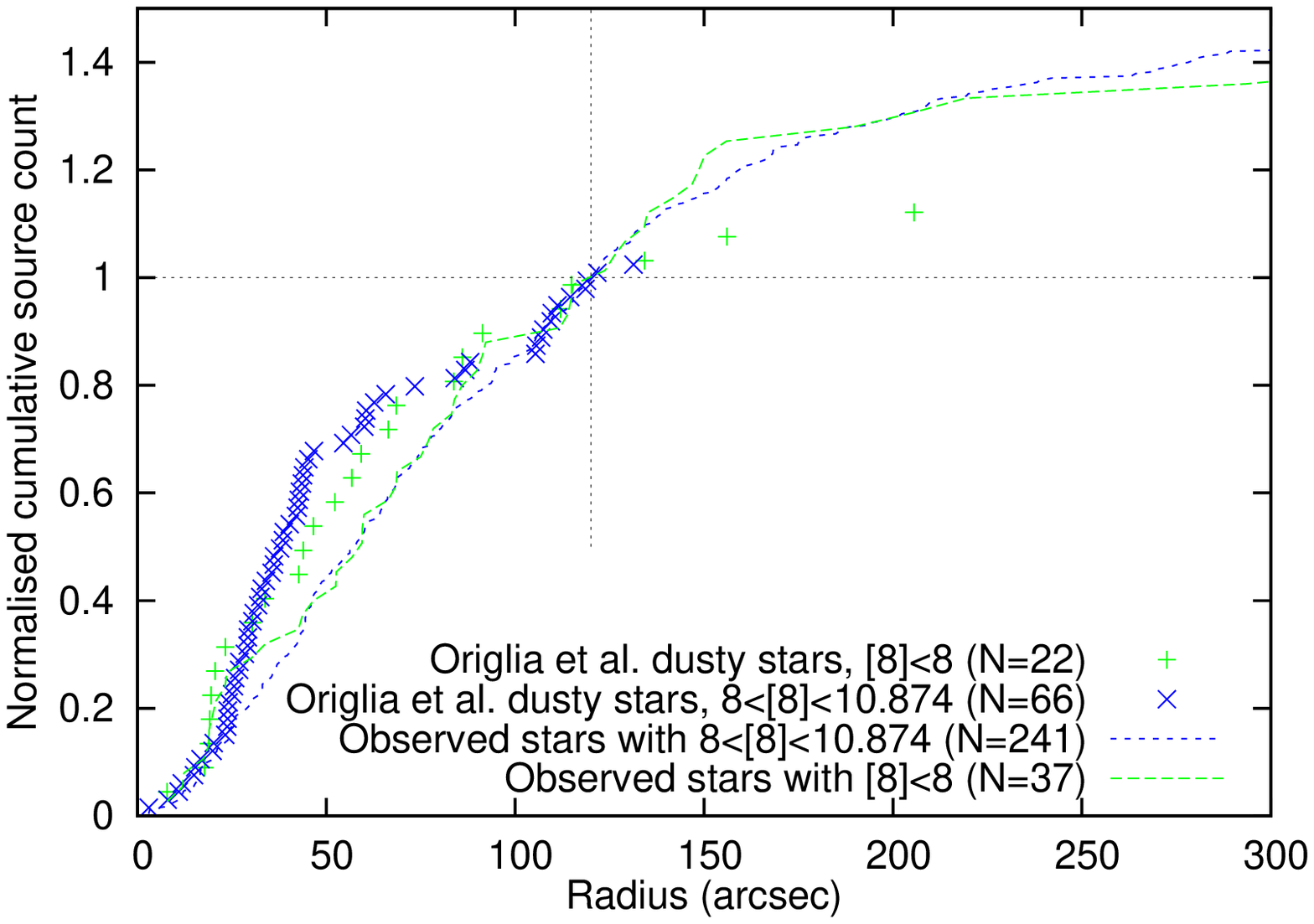}}
\caption{Radial distribution of bright and faint giants in our data, compared to the sample of bright and faint dusty candidates in the works of Origlia et al. All data are normalised to 2$^\prime$ and the number ($N$) of stars within that radius is shown for each sample.}
\label{OrigDensityFig}
\end{figure}

Of our remaining 120 objects, several exhibit a `step' jump of several tenths of a magnitude between $K_{s}$-band and 3.6 $\mu$m, but no additional excess at longer wavelengths. Two possibilities for this are large amounts of hot circumstellar dust, or bad cross-matching between optical/near-IR and \emph{Spitzer} data due to the very different resolutions between the two. Bad cross-matching is a problem that affects all surveys in crowded regions, but these objects would be retained under simple colour cuts in, say, ($K_{s}$--[8]).


We can separate bad cross-matching from real dust by examining where these objects lie in a colour-colour diagram (Figure \ref{RealDustFig}, top-left panel). Strictly speaking, this figure is a colour \emph{excess} -- colour \emph{excess} diagram, and thus in the following, we denote a colour excess in filter $Y$ with reference filter $X$ as ($X$--$Y$)$_{\rm e}$. This is equal to the observed colour minus the model atmosphere colour:
\begin{equation}
(X-Y)_{\rm observed} - (X-Y)_{\rm model}
\end{equation}
or, identically, the colour excess of the observation with respect to the model in filter $X$ minus the colour excess in filter $Y$:
\begin{equation}
(X_{\rm observed}-X_{\rm model}) - (Y_{\rm observed} - Y_{\rm model}).
\end{equation}
Two populations separate out in this diagram: ones with positive colours in both ($K_{s}$--[3.6])$_{\rm e}$ and ([3.6]--[8])$_{\rm e}$ and ones with positive colours in only ($K_{s}$--[3.6])$_{\rm e}$.

In Figure \ref{RealDustFig} (top-right panel), we have taken the SED of a typical star with ($K_{s}$--[3.6])$_{\rm e}$ $>$ 0 but ([3.6]--[8])$_{\rm e}$ $\approx$ 0: 2MASS\,00234588--7204488. We show circumstellar dust models made using metallic iron (optical constants from \citealt{OBA+88}) and amorphous carbon (optical constants from \citealt{ZMCB96}): two of the best grain species for producing flux at short wavelengths (e.g.\ \citealt{MSZ+10}). Neither can produce the observed SED as well as the naked photospheric model, multiplied upwards in flux. This is highly suggestive of source blending in the \emph{Spitzer} images, but not in the $K_{s}$-band data. Indeed, \citet{MAM06} report stars with \emph{HST} magnitudes of $F_{475W}$ = 14.65 and 15.17 mag at distances of 0.335$^{\prime\prime}$ and 1.74$^{\prime\prime}$ from this location, respectively, while \citet{SHO+07} reports the $K_{s}$-band magnitudes of the same two stars as 12.064 and 12.018 mag, respectively. On the other hand, 2MASS reports $K_{s}$ = 11.485 $\pm$ 0.073, which compares favorably to [3.6] = 11.539 $\pm$ 0.098 mag from \citet{BvLM+10}. It would therefore appear that this object is an unresolved blend in both 2MASS and \emph{Spitzer} data.

A third test we can do to ensure that these stars are unresolved blends is to look at the radial distribution of these objects (Figure \ref{RealDustFig}, bottom-left panel). Sources with significantly positive ([3.6]--[8])$_{\rm e}$ colours exhibit more central concentration than the main population of bright stars, falling off roughly as the square of projected source density, as might be expected from blending. Those with near-zero ([3.6]--[8])$_{\rm e}$ colours are even more centrally condensed, with the \emph{entire} group located within 2.5$^\prime$.

One can perform a similar test with the objects found in \citet{ORFF+07}. When doing so, we must bear in mind the completeness of our samples: comparing the distribution of their dusty sources to our total source distribution assumes both our datasets achieve the same completeness. The fact we work from the same data, however, would suggest that their completeness at 8 $\mu$m is not significantly different to ours. In absolute terms, false star tests retrieve 88\% of stars with [8] = 10.9 mag over the image, though this decreases to 75\% of stars between 1.5$^\prime$ and 2$^\prime$ and 58\% within 1$^\prime$. These percentages increase to 94\%, 88\% and 87\% for stars of [8] = 10.1 mag. We estimate that we should retrieve $\approx$90\% of stars of similar magnitude to Origlia et al.'s dusty stars (which have an average of [8] = 9.51 mag). Figure \ref{OrigDensityFig} shows that the radial distribution of both the bright ([8] $<$ 8 mag) and faint (8 $<$ [8] $<$ 10.874 mag) giants we detect in the original \citep{ORFF+07} \emph{Spitzer} images. 

The distribution of both the bright and faint giants follow each other within the Poissonian noise ($\sqrt{n}$). The distribution of bright dusty candidates in \citet{ORFF+07} matches the global distribution of bright giants we identify to within the Poissonian noise. The distribution of their faint dusty candidates, however, departs from the global distribution of stars with an identical magnitude range by over $2\sqrt{n}$, showing considerable central concentration. The lack of sources at radii $>2^\prime$ corroborates this finding: including stars over 2$^\prime$ from the cluster core increases the number of objects by $\approx$41\%, suggesting there should be $\approx$41\% (27) more faint dusty sources at large radii, whereas \citet{ORFF+07} only find two. Similarly, one expects $\approx$15 sources over 2.5$^\prime$ from the cluster core, whereas Origlia et al.\ find none. The differences we find in the radial distribution of their dusty objects and our total detected sample (with or without correction for unresolved targets) cannot be accounted for by differences in detection efficiency. A grossly centrally concentrated distribution of candidates therefore suggests that blending and its associated effects are a significant problem in the sample of \citet{ORFF+07}.

Finally, we compare the magnitudes for all stars which are common between the \citet{SHO+07} and 2MASS catalogues (Figure \ref{RealDustFig}, bottom-right panel). The majority of both groups of stars (with near-zero and with positive ([3.6]--[8])$_{\rm e}$ colours) scatter generally toward brighter 2MASS magnitudes. The lower-resolution of 2MASS therefore suggests that these stars are blends. We note that the brightest stars are not included in this test, as they are saturated in the data from \citet{SHO+07}.

To summarise, very warm dust does show an appreciable 3.6-$\mu$m excess (Figure \ref{RealDustFig}), hence a positive ($K_s$--[3.6]) colour. This would suggest that ($K_s$--[8]) is theoretically a better indicator of the presence of warm dust than ([3.6]--[8]). However, warm dust should still produce a substantial ([3.6]--[8]) colour, meaning it should not be greatly desensitised to the detection of warm dust when compared to ($K_s$--[3.6]). The above four observations indicate that the difference in resolution between the $K_s$ and [3.6] images has caused artificially-high ($K_s$--[3.6]) colours in both our data and that of \citet{ORF+10}, while ([3.6]--[8]) colours remain largely unaffected. We therefore conclude that ($K_{s}$--[8]) may detect warm circumstellar dust more efficiently than ([3.6]--[8]), but with three important caveats: (1) that in dense regions the resolution of the images used to determine colour should be broadly similar, to avoid unresolved blending affecting one band; (2) that stars of $\lesssim$3500 K will naturally show a ($K_s$--[3.6]) colour simply by virtue of being cool; and (3) neither colour can mitigate against the effect of blending with a red object, such as a background galaxy.

\subsubsection{Examination of individual stars}

With 120 targets remaining, it is now feasible to perform a visual inspection of each SED. We find that many faint stars have been selected merely because of their ($K_{s}$--[3.6]) colour, which suffers from the problems highlighted in \S\ref{K8Sect} and Figure \ref{RealDustFig}. This point is quite simple to address: we apply the colour cut shown in the upper-left panel of Figure \ref{RealDustFig}, to exclude stars with ([3.6]--[8])$_{\rm e}$ $<$0.1 mag which have no 24-$\mu$m data. This removes 40 of our 120 targets, including Origlia et al.'s ID-148, ID-181, ID-189, ID-296 and ID-309.

We now visually identify each of the remaining 80 targets on the \emph{Spitzer} 8-$\mu$m mosaic and remove those objects which lie on bands created by stars, unless the photometric excess observed is significantly greater than the flux in the band. We also remove targets lying on the first Airy ring of bright ($>$1000 L$_\odot$) stars, according to the same criterion, stars artificially brightened by bandwidth-effect ghosts, and stars on the edge of the image. Banding and blending on the Airy disc removes six stars due to V1 (including ID-222 and ID-373), five stars due to V8, two stars due to V4, two stars due to V27, one star due to LW1, one star due to LW7 and/or LW8, two stars due to LW9, three stars due to LW13 (one of which is also on the Airy disc of A19), and two stars due to LW18. In the cases of V27 and LW18, the effects of banding are compounded by banding from other bright stars in the cluster core. Two stars are here removed as their \emph{Spitzer} photometry is uncertain, as they lie on the edge of the mosaic. Selected objects which are actually banding-effect ghost images are also removed from around V21, LW3, LW9 and FBV45. This leaves 50 candidates, with the only faint stars in common with \citet{ORFF+07} being ID-67, ID-81 and ID-86.

Our remaining candidates now split into three main groups:
\begin{enumerate}
\item 29 bright ($>$1000 L$_\odot$) giants that are known variables and which (mostly) have known, spectroscopically-confirmed circumstellar dust \citep{vLMO+06,LPH+06}). We examine these in \S\ref{BrightXSSect}. 
\item 9 moderately-bright (250--1000 L$_\odot$) giants which are listed due to their positive ($K_{s}$--[3.6])$_{\rm e}$ and/or ([8]--[24])$_{\rm e}$ colours. We examine these immediately after this list.
\item 14 faint giants ($<$100 L$_\odot$) which have \emph{Spitzer} IRAC colours consistent with warm dust. We examine these in the remainder of this section.
\end{enumerate}

All nine moderately-bright giant stars lie in the crowded heart of the cluster. Eight of the nine stars (excluding F2, below) have not yet been rejected purely because their 24-$\mu$m photometry still suggested the presence of dust: i.e.\ they had a $>$10\% 24-$\mu$m excess. We note, however, that the PSFs of these stars are substantially more blended at 24 $\mu$m (FWHM = 6$^{\prime\prime}$) than at 8 $\mu$m (FWHM = 2$^{\prime\prime}$). Seven of the eight stars (excluding F1, below) share all the following criteria:
\begin{list}{\labelitemi}{\leftmargin=1em \itemsep=0pt}
\item positive ($K_{s}$--[3.6])$_{\rm e}$ and/or ([8]--[24])$_{\rm e}$ colours;
\item blended with several other fainter sources within the 24 $\mu$m PSF;
\item blended with much brighter objects outside the 24 $\mu$m PSF, including V1, LW10 and LW13 (only six objects, but the seventh only shows a 1.6$\sigma$ excess at 24-$\mu$m);
\item have (where available) $K_{s}$-band magnitudes which differ between the 2MASS and Salaris surveys, suggesting image resolution is affecting magnitude determination;
\item and no other evidence of circumstellar dust.
\end{list}

Given the problems caused by heavy blending which we have already highlighted in this section, we now remove these seven stars, which include ID-67, ID-81 and ID-86. This leaves 43 dusty candidates: 29 of these are above 1000 L$_\odot$, and 14 are below.

None of the 14 faint dusty candidates we find are in common with \citet{ORFF+07}. We therefore find it unlikely that they are dust enshrouded and are likely merely artifacts and outliers present in our own photometry: from our large number of \emph{Spitzer} sources (4462) we may expect outliers up to $\approx$3.6$\sigma$. To be rigorous, and because several of these targets lie outside of the field of view covered by \citet{ORFF+07}, we explore each target (F1--F14) individually:
\begin{list}{\labelitemi}{\leftmargin=1em \itemsep=0pt}
\item (F1) 00$^h$23$^m$52.8$^s$ --72$^\circ$04$^{\prime}$33$^{\prime\prime}$, 4305 K,  271 L$_\odot$: this star was selected purely on the basis of an 11\% excess at 24 $\mu$m, however the formal error on the flux is 23\%, meaning this is not a significant detection.
\item (F2) 00$^h$24$^m$05.8$^s$ --72$^\circ$04$^{\prime}$44$^{\prime\prime}$, 4058 K,  260 L$_\odot$: significant excess is present at 3.6 and 4.5 $\mu$m, but not detected at longer wavelengths. It is very close to a star four times its brightness, therefore the excess is likely a result of blend.
\item (F3) 00$^h$23$^m$47.7$^s$ --72$^\circ$04$^{\prime}$35$^{\prime\prime}$, 4939 K,  82 L$_\odot$: this object shows some flux excess at 5.8 $\mu$m, strong excess at 8$\mu$m, but is blended with stars of similar brightness at these wavelengths. It may also be affected by banding from V8, which is expected at this position but not visible on the images due to crowding.
\item (F4) 00$^h$25$^m$39.4$^s$ --72$^\circ$08$^{\prime}$21$^{\prime\prime}$, 4445 K,  73 L$_\odot$: identified by an 11\% (1.3--1.4$\sigma$) excess at both 5.8 and 8 $\mu$m, the low significance of this excess means this star is likely a spurious detection.
\item (F5) 00$^h$23$^m$54.1$^s$ --72$^\circ$04$^{\prime}$17$^{\prime\prime}$, 5210 K,  71 L$_\odot$: this star lies on the Airy rings of two slightly brighter stars and shows apparent excess at 5.8 and 8 $\mu$m as in F3 above. It is also poorly resolved from a nearby, slightly fainter companion, and has poor quality $J$-band photometry due to nearby companions. We suspect the apparent 29\% (3$\sigma$) excess is due to blending.
\item (F6) 00$^h$23$^m$59.8$^s$ --72$^\circ$05$^{\prime}$01$^{\prime\prime}$, 5265 K,  59 L$_\odot$: this star shows increasing excess at 3.6, 4.5 and 8$\mu$m. It is not detected at 5.8 $\mu$m. Poorly resolved from nearby bright stars in \emph{Spitzer} data, and with inconsistent $I$-band photometry, we suggest that this star shows excess due to blending.
\item (F7) 00$^h$24$^m$25.9$^s$ --72$^\circ$19$^{\prime}$37$^{\prime\prime}$, 5204 K,  51 L$_\odot$: flagged for investigation due to a 25\% (3$\sigma$) excess at 8 $\mu$m, photometry at other wavelengths shows considerable scatter above the reported errors. It may be a field star or unusual object, but the excess (if it exists) does not appear to be due to dust.
\item (F8) 00$^h$23$^m$39.2$^s$ --72$^\circ$04$^{\prime}$04$^{\prime\prime}$, 5185 K,  50 L$_\odot$: this star's PSF overlaps with the Airy rings of several nearby, unresolved, much brighter objects at 8 $\mu$m. It shows an excess at 3.6 and 8 $\mu$m, but a possible deficit at 4.5 $\mu$m. We consider the 18\% (3$\sigma$) excess to probably be due to blending.
\item (F9) 00$^h$24$^m$07.0$^s$ --72$^\circ$05$^{\prime}$46$^{\prime\prime}$, 4807 K,  47 L$_\odot$: we have no near-IR photometry for this star, meaning the photospheric contribution to the SED is less constrained. It was selected on the basis of a 16\% excess at 8 $\mu$m, however this is at a significance of $<$2$\sigma$, so is probably a spurious detection.
\item (F10) 00$^h$24$^m$10.0$^s$ --72$^\circ$03$^{\prime}$25$^{\prime\prime}$, 4765 K,  46 L$_\odot$: selected on the basis of a 34\% (3$\sigma$) excess at 8 $\mu$m, this star lies on the Airy ring of an unnamed star, which is 20$\times$ brighter at 8 $\mu$m, but not luminous enough (725 L$_\odot$) to be selected as a potential bright-star blend earlier in this section.
\item (F11) 00$^h$24$^m$11.4$^s$ --72$^\circ$04$^{\prime}$04$^{\prime\prime}$, 4983 K,  44 L$_\odot$: this star was flagged because of its ($K_{s}$--[3.6]) and ([5.8]--[8]) colours. We note that the same excess persists at 3.6, 4.5 and 8 $\mu$m, and that the $JHK_{s}$ fluxes are deficient compared to our model. The apparent excess in this object seems due to the differing resolutions at $K_{s}$ and 3.6 $\mu$m (see \S\ref{K8Sect}) and a spurious 5.8-$\mu$m flux.
\item (F12) 00$^h$26$^m$04.8$^s$ --72$^\circ$08$^{\prime}$16$^{\prime\prime}$, 5547 K,  42 L$_\odot$: identified due to a 34\% (2$\sigma$) excess at 8 $\mu$m, the excess in this object is likely only a spurious detection.
\item (F13) 00$^h$28$^m$35.1$^s$ --71$^\circ$51$^{\prime}$09$^{\prime\prime}$, 4010 K,  39 L$_\odot$: this star shows considerable excess at 5.8 and 8 $\mu$m (6$\sigma$ and 4$\sigma$, respectively). It is very cool for its luminosity, and is 23$^\prime$ from the cluster core. It seems most likely that this star is a luminous member of the SMC which was marginally brighter than our luminosity cutoff.
\item (F14) 00$^h$24$^m$31.7$^s$ --72$^\circ$06$^{\prime}$00$^{\prime\prime}$, 4576 K,  35 L$_\odot$: identified due to apparent excess at 3.6 and 5.8 $\mu$m (the only two \emph{Spitzer} detections), this star has photometry which departs significantly from the models at $IJHK_{s}$. The SED of this object is of insufficient quality and coverage to determine whether any IR excess truly exists.
\end{list}

In summary, of the above 14 stars, we find that the excess observed in eight is of low statistical significance ($< 3.6 \sigma$), that two have poor-quality SEDs, that three suffer from blending and that the final star is likely a member of the SMC. We remind the reader that, in any case, none of these stars was found to have mid-IR excess by \cite{ORF+10}. To conclude, we find no compelling evidence that any star in 47 Tuc produces dust until it reaches at least 1000 L$_\odot$. We find similar mid-IR excesses as those seen by \citet{ORFF+07}, but we find these in different objects. Such objects occur predominantly in heavily-blended environments, and are due to either image artifacts or under-reported photometric errors due to blending.

\subsubsection{Excess among the bright giants}
\label{BrightXSSect}

We now examine the bright giants which are left as candidates. From these, we remove LW20 and LW17, as they were flagged due to their variability, but have no mid-IR data with which to determine whether they are dusty. Identifying blended objects, we find that LW7 and LW8 are blended with each other and that V26 is blended with a star just over half its brightness. LW7 and LW8 are \emph{both} identified as having IR excess. It would appear likely that at least one of them harbours dust, however the amount of excess they show and the quality of their photometry prevents us from saying anything more substantial about any such dust. V26 is retained, as the star it is blended with is fainter, and because higher-resolution 8.6-$\mu$m photometry \citep{vLMO+06} confirms the excess seen by \emph{Spitzer}. Five more targets (LW3, FBV45, LW18, LW15 and LW6) were removed because their excesses are not statistically significant. The remaining 22 bright giants are listed in Table \ref{BrightXSTable}.

\begin{center}
\begin{table*}
\caption{Candidate dust-enshrouded stars above 1000 L$_\odot$ in 47 Tuc.}
\label{BrightXSTable}
\begin{tabular}{cccclll}
    \hline \hline
RA	& Dec	& Temp.	& Lum.		& Name$^1$& Dusty? & Notes$^2$\\
(J2000)	& (J2000)& (K)	& (L$_\odot$)	& \ 	& \   & \ \\
    \hline
00 24 12.64 & --72 06 39.7 & 3623 & 4824 & V1   & Y & Silicate dust (see vLMO, LPH)\\
00 24 08.57 & --72 03 54.6 & 3578 & 3583 & V8   & Y & Silicate dust (see vLMO, LPH); x06 in MvL\\
00 24 18.55 & --72 07 58.8 & 3738 & 3031 & V2   & Y & Silicate dust (see vLMO, LPH)\\
00 25 15.94 & --72 03 54.7 & 3153 & 2975 & V3   & Y & Excess in vLMO, LPH; x08 in MvL\\
00 24 00.50 & --72 07 26.7 & 3521 & 2603 & V4   & Y & Silicate \& oxide dust (see LPH), $\sim$RGB-tip\\
00 24 07.71 & --72 04 31.4 & 3500 & 2541 & V26  & Y & x07 in MvL, =LW13\\
00 24 02.46 & --72 05 07.2 & 3543 & 2324 & LW10 & Y & x01 in MvL, also in vLMO\\
00 23 50.35 & --72 05 50.3 & 3575 & 2301 & V21  & Y & Oxide dusts (see LPH)\\
00 23 58.15 & --72 05 49.1 & 3374 & 2204 & LW9  & Y & \\
00 24 15.12 & --72 04 36.3 & 3510 & 2140 & V27  & Y & \\
00 23 29.98 & --72 22 36.3 & 3565 & 2122 & Lee1424& Y & \\
00 24 21.70 & --72 04 13.1 & 3526 & 2096 & A19  & Y & \\
00 24 03.97 & --72 05 09.9 & 3713 & 2079 & LW12 & Y & \\
00 24 09.39 & --72 04 48.8 & 3816 & 1640 & MVx3 & Y & x03 in MvL\\
00 24 23.15 & --72 04 22.8 & 3738 & 1638 & LW19 & ? & x05 in MvL\\
00 24 14.44 & --72 05 09.1 & 3602 & 1575 & V20  & Y & \\
00 24 08.88 & --72 02 59.5 & 3684 & 1528 & V22  & ? & JHK photometry questionable\\
00 24 25.67 & --72 06 29.9 & 3763 & 1374 & V6   & ? & No excess in LPH\\
00 25 03.60 & --72 09 31.7 & 3741 & 1363 & V5   & ? & No excess in LPH\\
00 25 09.15 & --72 02 39.6 & 3692 & 1297 & V18  & Y & Silicate dust (see vLMO, LPH)\\
00 24 29.51 & --72 09 07.5 & 3775 & 1259 & V23  & ? & \\
00 22 58.41 & --72 06 56.1 & 3657 & 1029 & V13  & Y & Oxide dusts (see LPH)\\
    \hline
\multicolumn{7}{p{0.9\textwidth}}{$^1$Variable (V, LW, A) names from \citet{Clement97}; and \citet{LW05}. Other names from: Lee --- \citet{Lee77}; FBV --- \citet{FBV+02}; MV --- \citet{MvL07}.}\\
\multicolumn{7}{p{0.9\textwidth}}{$^2$vLMO --- \protect\citet{vLMO+06}; LPH --- \protect\citet{LPH+06}; MvL --- \protect\citet{MvL07}.}\\
    \hline
\end{tabular}
\end{table*}
\end{center}

We again examine the SED and literature data of each star in turn. We include in these SEDs literature mid-IR photometry from \citet{OSA+97} (ESO-3.6m/TIMMI); \citet{OFFPR02} (\emph{ISO}); \citet{ITM+07} (\emph{AKARI}) and the \emph{AKARI} point-source catalogue \citep{KAC+10}. The \emph{ISO} and \emph{AKARI} data are taken at coarser resolution than that provided by \emph{Spitzer} IRAC, so caution must be taken when dealing with stars near the cluster core. We also include photometric data and spectra from \citet{vLMO+06} (ESO-3.6m/TIMMI2). Comparison with the \emph{Spitzer} IRS data of \citet{LPH+06} allows us to more-precisely determine dust compositions.

In the absence of spectroscopy, we have seen that identification of circumstellar dust production based solely on \emph{Spitzer} photometry can be problematic. Strong IR excess which increases in magnitude (but not necessarily flux) with wavelength is generally a reliable indicator of the presence of circumstellar dust. Spectra are needed to conclusively identify the dust composition, though the presence or absence of silicates or other oxides can sometimes be determined in comparatively-isolated stars with excess between 8 and 24 $\mu$m. Based on the quality of the SEDs, literature observations of these stars, and the IR excess we observe, we indicate in Table \ref{BrightXSTable} whether a star is clearly (Y) or probably (?) dusty. Further analysis of the dust minerology and mass-loss rates of these stars can be found in the accompanying Paper IV.


\section{Discussion}
\label{SectDisc}

\subsection{RGB mass loss and metallicity}

Modelling of horizontal branch star masses (\S\ref{SectIsos}) implies that integrated RGB mass loss per star in 47 Tuc does not greatly exceed that in $\omega$ Cen (see also Paper I). This is despite the clusters being of similar age, and $\omega$ Cen being (on average) almost a factor of ten more metal-poor. Similar HB stellar masses and integrated RGB mass-loss rates in the two clusters would imply that the dominant RGB mass loss process is metallicity-\emph{independent} for globular cluster stars.

This would corroborate findings in other clusters (open and globular, including NGC 6791 and $\omega$ Cen) that the super-solar metallicity open cluster NGC 6791 that show RGB and AGB dust production is similar to solar-metallicity expectations and that there is no unexpected absence of RGB stars due to super-solar mass loss \citep{vLvLS+07,vLBM08}. This is not to say that dust production is necessarily metallicity-independent: lower-metallicity stars are warmer at a given luminosity, therefore their chromospheric mass-loss rates are expected to be higher than their higher-metallicity counterparts \citep{SC05}. Better estimation of horizontal branch star masses are required in order to improve models before this can be fully substantiated, however.

\subsection{RGB dust formation}

We find no reliable evidence that dust is being produced by any RGB star below 1000 L$_\odot$. This is concurrent with the recent findings of \citet{BvLM+10}. We do not confirm the claims of dust production in lower-luminosity RGB stars made in \citet{ORFF+07,ORF+10}. Our two studies use the same \emph{Spitzer} data, but differ in the software used to determine photometric magnitudes. Given the strong artifacts present in the original data (Figure \ref{BandingFig}), we would advise care in any claim of excess among fainter stars in regions affected by these artifacts (including the cluster core). Higher-resolution data, such as those of Momany et al.\ (in prep.), are needed before the claim of dust production at low luminosities can be conclusively proven or refuted, and we await their results with anticipation.

We instead find that dusty mass loss begins in 47 Tuc at $\sim$1000 L$_\odot$ and becomes commonplace by $\sim$2000 L$_\odot$. Between 1000 and 2000 L$_\odot$, it is unclear why dust production is not occurring around all stars. In this region, nine out of 56 show possible IR excess, while three (x03, V18 and V13) show strong mass loss. This could represent episodic dust production on either short or long timescales, potentially related to pulsation. Alternatively, it could represent a difference in dust production between RGB and AGB stars (due to, say, gravity or dredge up), which we cannot separate at these luminosities. The luminosity at which dust production first occurs in 47 Tuc is therefore very similar to those we find in $\omega$ Cen (Paper I).


\section{Conclusions}
\label{SectConc}

In this work, we have used spectral energy distributions to establish physical parameters for stars in the globular cluster 47 Tuc. We have used these to investigate the basic parameters of the cluster as a whole, and the later stages of the evolution of its stars. We summarise our conclusions as follows:
\begin{list}{\labelitemi}{\leftmargin=1em \itemsep=0pt}
\item Simple isochrone fits to the cluster's Hertzsprung--Russell diagram corroborate the established distance and age of the cluster. We find $d = 4611 ^{+213}_{-200}$ pc and $t = 12 \pm 1$ Gyr.
\item HB models show that mass loss on the RGB is unlikely to greatly exceed that in the similarly-aged but much more metal-poor cluster $\omega$ Cen, implying that RGB mass loss does not vary with metallicity. We do not rule out any correlation of dust production with metallicity: our results apply only to integrated mass loss.
\item We find that apparent IR excess in stars below $L = 1000$ L$_\odot$ is almost certainly due to artifacts in the original \emph{Spitzer} maps and under-reported photometric errors caused by blending, a finding contrary to the results of \citet{ORFF+07} and \citet{ORF+10}.
\item Some stars above $L = 1000$ L$_\odot$ show IR excess consistent with dust production. The fraction of dust-producing stars approaches unity above $L \approx 2000$ L$_\odot$.
\end{list}

\vspace{5 mm}
\noindent
{\bf Acknowledgments:} We are grateful to Livia Origlia for her co-operation in the production of this paper, for her contribution to the analysis and for sharing her original data with us. This paper uses observations made using the \emph{Spitzer Space Telescope} (operated by JPL, California Institute of Technology under NASA contract 1407 and supported by NASA through JPL (contract number 1257184)); observations using \emph{AKARI}, a JAXA project with the participation of ESA; and data products from the Two Microns All Sky Survey, which is a joint project of the University of Massachusetts and IPAC/CIT, funded by NASA and the NSF.


\end{document}